\documentclass[twocolumn]{aastex631}

\newcommand{\dmunits}{pc~cm$^{-3}$}

\newcommand{\diskfrb}{FRB~20230930A\,}
\newcommand{\halofrb}{FRB~20230506C\,}
\newcommand{\realfast}{\textit{realfast~}}
\newcommand{\hi}{\ion{H}{1}}
\newcommand{\kms}{km s$^{-1}$}

\usepackage{amsmath}
\usepackage{todonotes}
\usepackage{graphicx}
\begin{document}

\title{Evidence for a hot galactic halo around the Andromeda Galaxy using fast radio bursts along two sightlines.}

\shorttitle{FRBs probing M31's halo}
\shortauthors{R. Anna-Thomas et al.}

\author[0000-0001-8057-0633]{Reshma Anna-Thomas}
\affiliation{West Virginia University, Department of Physics and Astronomy, P. O. Box 6315, Morgantown, WV, USA}
\affiliation{Center for Gravitational Waves and Cosmology, West Virginia University, Chestnut Ridge Research Building, Morgantown, WV, USA}
\affiliation{ASTRON, Netherlands Institute for Radio Astronomy, Oude Hoogeveensedĳk 4, 7991 PD Dwingeloo, The Netherlands}
\affiliation{Anton Pannekoek Institute for Astronomy, University of Amsterdam, Science Park 904, 1098 XH Amsterdam, The Netherlands}

\correspondingauthor{Reshma Anna-Thomas}
\email{thomas@astron.nl}

\author[0000-0002-4119-9963]{Casey J. Law}
\affiliation{Cahill Center for Astronomy and Astrophysics, MC 249-17 California Institute of Technology, Pasadena CA 91125, USA}
\affiliation{Owens Valley Radio Observatory, California Institute of Technology, Big Pine CA 93513, USA}

\author[0000-0001-9605-780X]{Eric W. Koch}
\affiliation{Center for Astrophysics $\mid$ Harvard \& Smithsonian, 60 Garden St., 02138 Cambridge, MA, USA}

\author[0000-0002-5025-4645]{Alexa~C.~Gordon}
\affiliation{Center for Interdisciplinary Exploration and Research in Astrophysics (CIERA) and Department of Physics and Astronomy, Northwestern University, Evanston, IL 60208, USA}

\author[0000-0002-4477-3625]{Kritti Sharma}
\affiliation{Cahill Center for Astronomy and Astrophysics, MC 249-17 California Institute of Technology, Pasadena CA 91125, USA}
\affiliation{Owens Valley Radio Observatory, California Institute of Technology, Big Pine CA 93513, USA}

\author[0000-0002-7502-0597]{Benjamin F. Williams}
\affiliation{Astronomy Department, University of Washington, Seattle, WA, 98195}

\author[0000-0001-9504-7386]{Nickolas~M.~Pingel}
\affiliation{University of Wisconsin–Madison, Department of Astronomy, 475 N Charter St, Madison, WI 53703, USA}

\author[0000-0003-4052-7838]{Sarah Burke-Spolaor}
\affiliation{West Virginia University, Department of Physics and Astronomy, P. O. Box 6315, Morgantown, WV, USA}
\affiliation{Center for Gravitational Waves and Cosmology, West Virginia University, Chestnut Ridge Research Building, Morgantown, WV, USA}
\affiliation{Sloan Fellow}

\author[0000-0002-3038-3896]{Zhuo Chen}
\affiliation{Astronomy Department, University of Washington, Seattle, WA, 98195} 

\author{Jordan Stanley}
\affiliation{West Virginia University, Department of Physics and Astronomy, P. O. Box 6315, Morgantown, WV, USA}
\affiliation{Center for Gravitational Waves and Cosmology, West Virginia University, Chestnut Ridge Research Building, Morgantown, WV, USA}

\author{Calvin Dear}
\affiliation{West Virginia University, Department of Physics and Astronomy, P. O. Box 6315, Morgantown, WV, USA}
\affiliation{Center for Gravitational Waves and Cosmology, West Virginia University, Chestnut Ridge Research Building, Morgantown, WV, USA}

\author{Frank Verdi}
\affiliation{Cahill Center for Astronomy and Astrophysics, MC 249-17 California Institute of Technology, Pasadena CA 91125, USA}
\affiliation{Owens Valley Radio Observatory, California Institute of Technology, Big Pine CA 93513, USA}

\author[0000-0002-7738-6875]{J. Xavier Prochaska}
\affiliation{Department of Astronomy and Astrophysics, University of California, Santa Cruz, CA 95064, USA}
\affiliation{Kavli Institute for the Physics and Mathematics of the Universe, 5-1-5 Kashiwanoha, Kashiwa 277-8583, Japan}
\affiliation{Division of Science, National Astronomical Observatory of Japan, 2-21-1 Osawa, Mitaka, Tokyo 181-8588, Japan}

\author[0000-0003-4056-9982]{Geoffrey C. Bower}
\affiliation{Academia Sinica Institute of Astronomy and Astrophysics, 645 N. A'ohoku Pl., Hilo, HI 96720, USA}
\affiliation{Department of Physics and Astronomy, University of Hawaii at Manoa, 2505 Correa Road, Honolulu, HI 96822, USA}

\author{Laura Chomiuk}
\affiliation{Center for Data Intensive and Time Domain Astronomy, Department of Physics and Astronomy, Michigan State University, East Lansing, MI 48824, USA}

\author[0000-0002-7587-6352]{Liam Connor}
\affiliation{Center for Astrophysics $\mid$ Harvard \& Smithsonian, 60 Garden St., 02138 Cambridge, MA, USA}

\author[0000-0002-6664-965X]{Paul B.\ Demorest}
\affiliation{National Radio Astronomy Observatory, P.O. Box O, Socorro, NM 87801, USA}

\author[0000-0002-7374-935X]{Wen-Fai Fong}
\affiliation{Center for Interdisciplinary Exploration and Research in Astrophysics (CIERA) and Department of Physics and Astronomy, Northwestern University, Evanston, IL 60208, USA}

\author[0000-0002-2028-9329]{Anya Nugent}
\affiliation{Center for Interdisciplinary Exploration and Research in Astrophysics (CIERA) and Department of Physics and Astronomy, Northwestern University, Evanston, IL 60208, USA}
\affiliation{Center for Astrophysics $\mid$ Harvard \& Smithsonian, 60 Garden St., 02138 Cambridge, MA, USA}

\author[0000-0003-4793-7880]{Fabian Walter}
\affiliation{{Max Planck Institut f\"ur Astronomie, K\"onigstuhl 17, D-69117, Heidelberg, Germany}}

\begin{abstract}
Fast Radio Bursts (FRBs) are millisecond-duration radio transients that serve as unique probes of ionized extragalactic matter. We report the discovery and localization of two FRBs piercing the Andromeda Galaxy (M31) with the \realfast transient-detection system at the Very Large Array. These unique sightlines enable constraints on M31’s electron density distribution. We localized \diskfrb\ to a host galaxy at redshift $z=0.0925$ and \halofrb\ to a host galaxy at redshift $z=0.3896$. After accounting for the dispersion contributions from the Milky Way, the host galaxies, and the intergalactic medium, we estimate M31’s contribution to be $26–239$ \dmunits\ toward \diskfrb\ and $51–366$ \dmunits\ toward \halofrb, within  the 90\% credible interval (CI). By modeling the M31 disk’s contribution, we isolate the halo component and find that M31’s halo contributes $7–169$ \dmunits\ along \diskfrb\ (90\% CI). The inferred values of $\rm DM_{M31,halo}$ from the FRBs are consistent with predictions from a modified Navarro–Frenk–White (mNFW) profile at the corresponding impact parameter. The cool and warm phase gas is unlikely to account for the $\rm DM_{M31,halo}$ unless the ionization fraction is as high as 90\%. While limited to two sightlines, these results offer tentative evidence for the existence of a hot halo surrounding M31. We also discuss the potential contribution of other foreground structures, particularly in explaining the DM excess observed in \halofrb. This work demonstrates how FRBs can be used to probe the circumgalactic medium of intervening galaxies.
\end{abstract}
\section{Introduction}
A hot corona around our Galaxy was first predicted by \cite{Spitzer1956} as the cause of absorption lines in the spectra of stars at high Galactic latitude. The exploration of this circumgalactic gas was then carried out by absorption line spectroscopy of bright background sources. The circumgalactic medium (CGM) regulates the inflow and outflow of gases and therefore plays an important role in galaxy evolution. The CGM represents a multiphase metal-enriched gas reservoir around all galaxies and likely extends beyond the virial radius, $R_{\rm vir}$ \citep[e.g.][]{Tumlinson2017}. Apart from absorption line studies, independent constraints on the CGM were measured in the microwave regime by studying the distortion in the cosmic microwave background spectrum by the hot electrons in the halo. This process termed as the thermal Sunyaev- Zeldovich (SZ) effect \citep{SZ}, has been used to study the CGM of nearby galaxies \citep{Bregman2022}.  X-ray emission due to thermal bremsstrahlung \citep{Li2018} has also been used to study the hot CGM around nearby galaxies.

The Andromeda galaxy (M31), at a distance of $761\pm11$ kpc \citep{Li-M312021}, is the closest large galaxy to the Milky Way and its halo subtends an angle of 30$^\circ$ in the sky, making it a perfect candidate to study the CGM. \cite{Lehner2015} discovered evidence of a massive extended CGM around M31 using far-ultraviolet absorption lines of metal ions corresponding to a cold phase ($\rm T\leq10^4~K$) and a warm phase ($\rm T\sim 10^{5.5}~K$). They showed that this CGM is bound, exists in multiple phases, and its ionization fraction increases with the radius from the center. The baryon mass within $R_{\rm vir}$ for the cold-warm phase ($\rm T\sim 10^4 - 10^{5.5}~K$) CGM was estimated to be ${\rm > 4 \times 10^{10}}(Z/0.3Z_{\odot})^{-1}~\rm M_{\odot}$ \citep{Lehner2020}, where $Z$ denotes the metallicity of the medium. Observational evidence of the hot ($\rm T\ge10^6~K$) phase of the CGM, presumably surrounding the cold phase, is still lacking.

The frequency-dependent dispersion associated with fast radio bursts (FRBs; \cite{Lorimer2007}) provides a unique probe to measure the baryon content of the intervening medium between the source and the observer. The dispersion measure, ${\rm DM} = \int_d n_e dl$ is a direct measurement of the electron density in the observer's line of sight. Previous studies have used FRBs to measure the electron density of the halos of intervening galaxies \citep{Prochaska2019b, Wu&McQuinn2023}, including the Milky Way \citep{Ravi2023, Cook2023}, the intracluster medium \citep{Connor2023}, and the intergalactic medium \citep{Macquart2020}. Previously, \cite{Connor2020, vanLeeuwen2023} detected FRBs skewering the M31-M33 halos and suggested that the shared plasma of the group contributed to the DM of the FRB. In a different study using hundreds of FRBs detected by the Canadian Hydrogen Intensity Mapping Experiment (CHIME), \cite{Connor2022} finds weak evidence for DM excess contributed by the halos of M31 and M33, whereas \cite{Wu&McQuinn2023} found the evidence to be only marginal. Therefore, it is crucial to detect more FRBs intersecting the local group galaxies to reliably understand the CGM around these galaxies and its potential impact on FRB detection.  

\realfast \citep{Law2015,Law2018} is a real-time commensal transient search, detection, and localization system operating at the Karl J. Jansky Very Large Array (VLA) between the frequencies 1 to 10 GHz. 
\realfast has been instrumental in the localization of many FRBs like the first repeating FRB 20121102A \citep{Chatterjee2017}, FRB 20180916B \citep{Aggarwal2020} and FRB 20190520B \citep{Niu2022}. It has also discovered FRB 20190614D \citep{Law2020} and a Galactic pulsar-like source J1818--1531 \citep{Anna-Thomas2024}. In this paper, we discuss the discovery and localization of two FRBs by \realfast that pierce through M31.

This paper is organized into multiple sections: \S\ref{sec:obs} describes radio and optical observations and data reduction, \S\ref{sec:dm_budget} discusses the DM budget and constraints the DM contribution from M31, \S\ref{sec:discussion} discusses various DM contributions, and \S\ref{sec:conclusion} summarizes the results. Throughout this paper, we adopt
cosmological model and parameters from the Planck 2018  analysis for any calculation that requires assumed cosmology \cite{Planck2018}.

\begin{figure*}
\centering\includegraphics[width=1.0\textwidth]{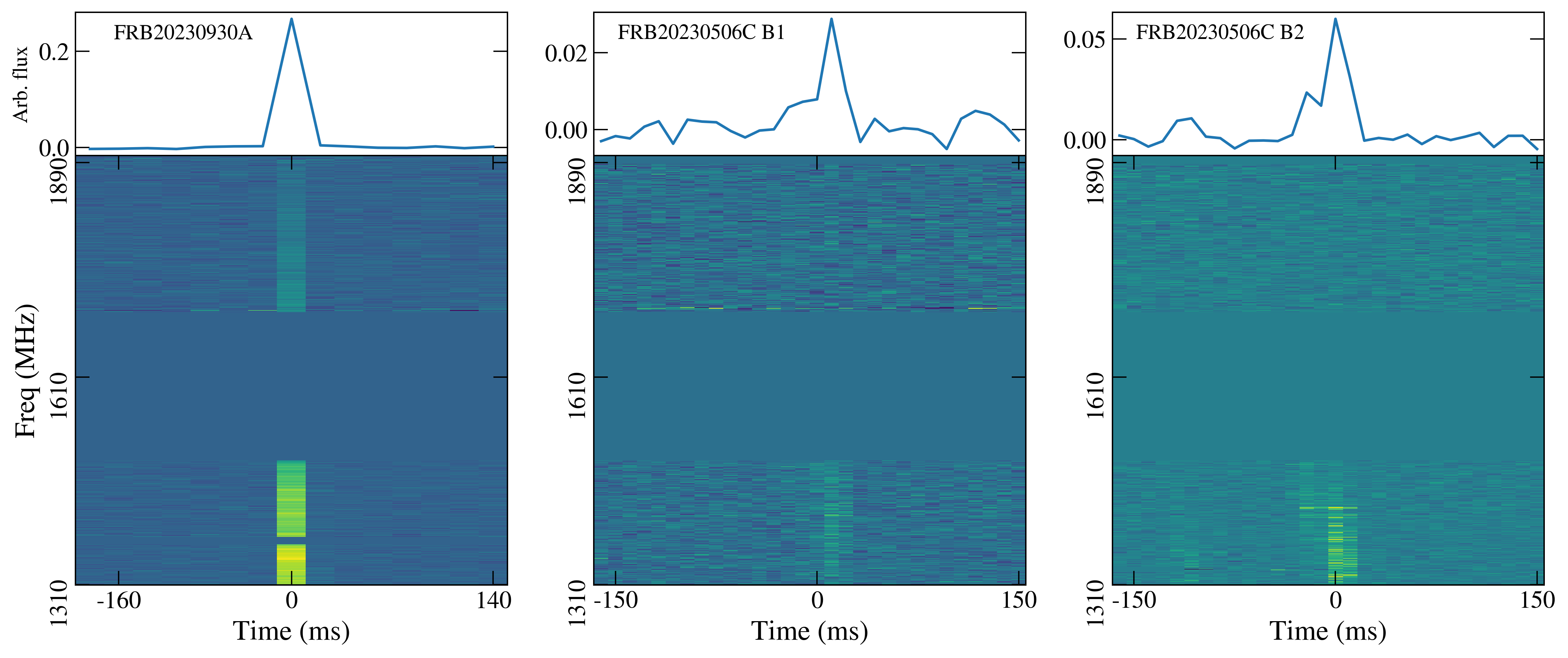}
    \caption{Dynamic spectrogram (bottom) and frequency‑averaged time profile (top) of
the \realfast\ bursts, dedispersed at their detection DM. The data have 10 ms time
resolution and 1 MHz frequency resolution; note that channels between 1490 MHz and
1690 MHz were not recorded by the \realfast\ system}
    \label{fig:spectrogram}
\end{figure*}
\section{Observations and data reduction} \label{sec:obs}
\subsection{VLA/Realfast observation}
The VLA was observing the Local Group Legacy survey (EVLA 20A-346; P.I. Adam Leroy) which targets deep, high spatial resolution imaging of 21 cm and L-Band continuum emission of six local group galaxies, including M31. The \realfast transient-detection system ran commensally with these observations, ingesting correlated voltages sampled at 10 ms resolution. The software \textsc{rfpipe} applies online calibration, searches for bursts at many different trial DM and widths. Candidates above $8\sigma$ fluence limit for a 10 ms image ($0.29$ Jy ms at L-band) triggered the recording of a few seconds of visibility data centered around the candidate. These candidates are then visually inspected by the \realfast users. 

On May 6, 2023 (MJD 60070) and on September 30, 2023 (MJD 60217), \realfast detected two FRBs, \diskfrb\ and \halofrb, when the telescope was pointed at J2000  RA = $\rm 00h41m40.844s$, DEC = $\rm 41^\circ44$\arcmin$02.379$\arcsec (field=M31LARGE\_47) and RA = $\rm 00h48m42.535s, DEC=42^\circ00$\arcmin$51.122$\arcsec (M31LARGE\_4), respectively. This triggered the download of 10-ms Science Data Model (SDM) data. Repeat bursts from \halofrb\ were subsequently detected on August 26, 2023 (MJD 60182) and September 28, 2023 (MJD 60215). The VLA was in the B configuration on 60070, and in A configuration on 60182, 60215, and 60217. The frequency range of the \realfast data is 1.308 -- 2.012 GHz and is divided into 8 spectral windows, each having 64 channels with 1 MHz resolution. We also note that the frequencies between 1.49 --1.69 GHz were not recorded due to the spectral set up of the primary observation. The \realfast system also did not store the visibilities on MJD 60182 for \halofrb\ and it was only detected by the realtime pipeline. For both fields, we observed the sources J2355+4950 for phase calibration at regular intervals, 3C48 for flux calibration and J1800+7828 for polarization calibration  at the end of each track. 

The burst profiles and spectrograms are shown in Figure \ref{fig:spectrogram}. 
We used the package \textsc{Burstfit} on the \realfast bursts to do spectro-temporal modeling of the bursts. We modeled the temporal profile of the burst using a Gaussian convolved with an exponential tail and the time-averaged spectra of the burst using a simple Gaussian as done in \cite{Aggarwal2021}. We do not report the scattering timescale ($\tau$) since the ratio of $\tau/\sigma_t < 3$, which indicates that the scattering is not significant. Here $\sigma_t$ represents the standard deviation of the Gaussian pulse. Using this modeling, we fit for the width, DM, center frequency and the bandwidths of each bursts.  

The S/N maximized DM of \diskfrb\ is $456^{+0.5}_{-0.6}$ \dmunits\ and the DM of the brightest burst of \halofrb\ is $772^{+3}_{-2}$ \dmunits . The burst properties of the three \realfast bursts are given in Table \ref{tab:burst_properties}.

\begin{table*}
    \centering

        \begin{tabular}{cccc}
        
        \hline
            \textbf{Properties} & \textbf{\diskfrb} &\textbf{ \halofrb B1} & \textbf{\halofrb B2}\\
            \hline
            RA & $\rm 00h42m01.734s$  &  $\rm 00h48m23.9579s$& $\rm 00h48m23.9608s$\\
            $\rm \Delta RA$& 0.1\arcsec & 0.12\arcsec &0.12\arcsec \\
            DEC& $+41^\circ25\arcmin02.4143\arcsec$ & $+42^\circ00\arcmin21.8822\arcsec$ & $+42^\circ00\arcmin21.9249\arcsec$ \\
            $\rm \Delta Dec$&  0.18\arcsec&0.18\arcsec  &0.18\arcsec \\
            S/N& 55 & 14 &38 \\
            MJD& 60217.2074113 & 60070.7238837 & 60215.2046195\\
            DM ($\rm pc~cm^{-3}$)& $456^{+0.5}_{-0.6}$ & $761^{+5}_{-5}$ & $772^{+3}_{-2}$\\
            Flux (Jy)& $0.27\pm0.004$ &  $0.05\pm0.001$ & $0.14\pm0.002$\\
            Width (ms)& $8.7^{+1.1}_{-1.1}$& $18^{+1.4}_{-1.1}$& $17^{0.7}_{-0.7}$  \\
            $\mu_{f}$ (MHz)& $1329^{+10}_{-10}$ & $1392^{+5}_{-5}$ & $1351^{+3}_{-3}$\\
            $\sigma_{f}$ (MHz)& $134^{+7}_{-7}$ & $59^{+6}_{-5}$ & $58^{+3}_{-3}$ \\
            \hline
        \end{tabular}
    \caption{Observed properties of all \realfast\ bursts.\\
    S/N is the image plane signal-to-noise obtained during the offline refinement of the bursts.\\
    MJD is the time of arrival of the bursts corrected to the barycentric frame of reference (TDB) and infinite frequency.\\
    DM is the S/N maximizing dispersion measure obtained from \textsc{burstfit}.\\
    Flux as obtained from CASA's \texttt{imfit}.\\
    Width of the burst in ms as obtained from \textsc{burstfit}.\\
    $\mu_f$ is the center frequency of the burst obtained from  \textsc{burstfit}.\\
    $\sigma_f$ is the bandwidth of the burst spectra from \textsc{burstfit}.\\
    NOTE: \realfast did not record the data for a burst of \halofrb\ on MJD 60182 and hence it is omitted here.}
    \label{tab:burst_properties}
\end{table*}



\subsection{Realfast imaging and localization}
The real-time images of the candidates are convolved with the point spread function, and are made with several assumptions like coarse DM grid, non-optimal image size, simpler calibration model, etc. To rectify this, we followed the steps in \cite{Anna-Thomas2024} for post processing and offline imaging of the \realfast data. An additional step of spectral‑window mapping was applied to ensure that solutions were applied to the correct frequencies. As mentioned earlier, \realfast did not record the fast-sampled visibility data for the FRB detected on MJD 60182. For the other three bursts, we created cleaned images using \textsc{CASA} and fitted the bursts by a 2D elliptical Gaussian using \texttt{imfit} to get the flux density, centroid position, and $1\sigma$ image-plane uncertainties. The statistical uncertainty in the position of \diskfrb\ is $\rm \Delta RA_{stat} = 0.01$\arcsec, $\rm \Delta Dec_{stat}=0.01$\arcsec and for \halofrb\ is $\rm \Delta RA_{stat} = 0.01$\arcsec, $\rm \Delta Dec_{stat}=0.007$\arcsec.
\begin{figure*}
    \centering
    \includegraphics[width=1.0\textwidth]{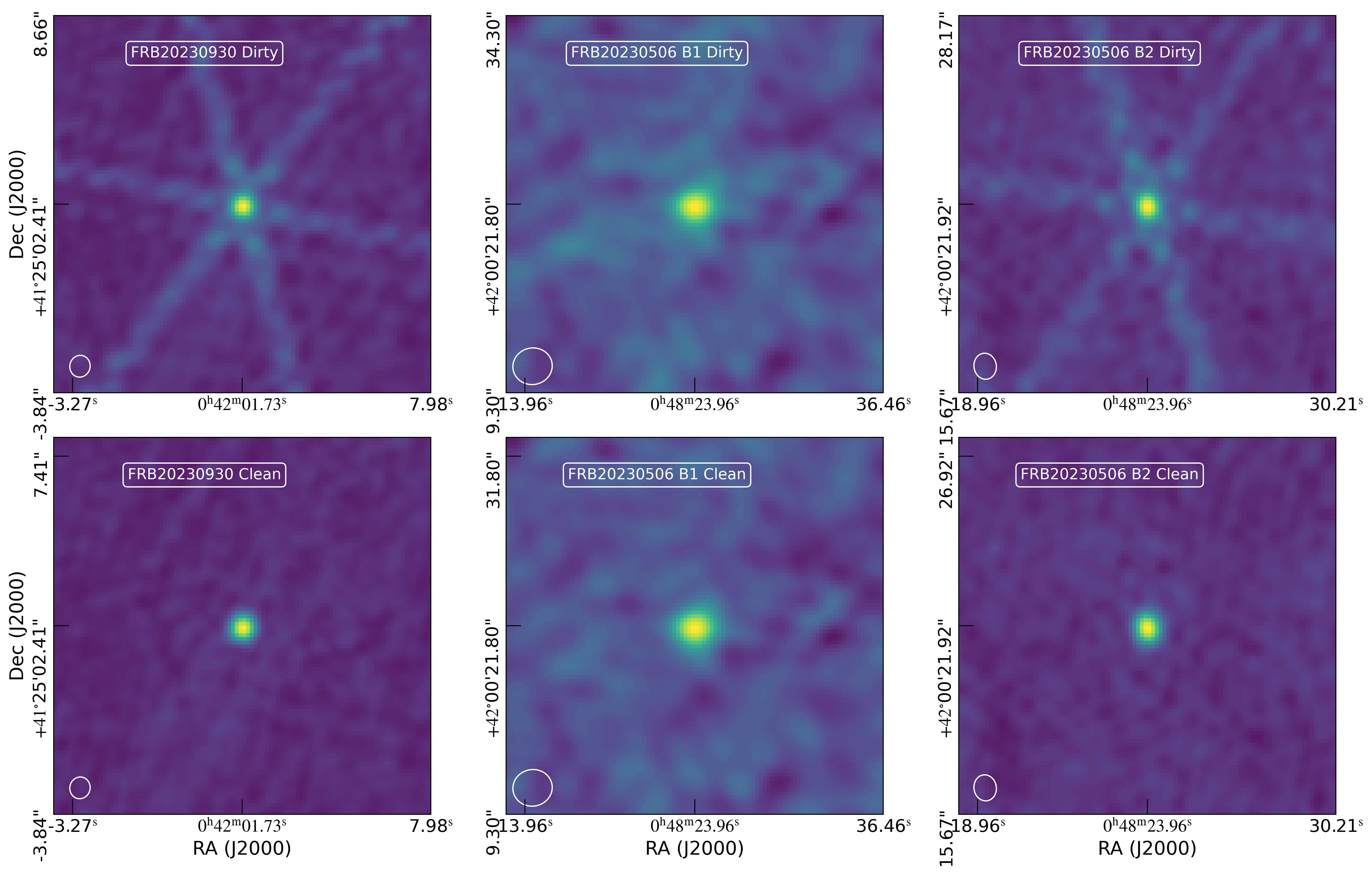}
    \caption{The dirty and clean maps of the \realfast localization of the bursts. \diskfrb\ and \halofrb\ B2 were observed in VLA A-configuration, and the \halofrb\ B1 was observed in VLA B-configuration.}
    \label{fig:cleandirty}
\end{figure*}

To determine the systematic offsets in the burst positions, we made a deep image of the VLA pointing of the fields of \diskfrb\ and \halofrb. For the \halofrb, we used the scan on the MJD of its brightest burst (B2). We ran a source extraction software \textsc{PyBDSF} \citep{pybdsf} and identified 75 radio sources in the field. We then selected only the bright, compact sources using the criteria: 1) The peak intensity per beam of the source (in Jy/beam) should be 0.7 times greater than the total integrated flux density of the source (in Jy) in 1.5 GHz images, 2) the S/N of the source (ratio of peak intensity and the root-mean-square of the background) should be greater than 5. We were left with 27 sources in the fields of both FRBs. We cross-matched the positions of these radio sources with optical PAN-STARRS DR2 catalog, which is referenced to GAIA2 astrometric reference frame. We then subtracted the coordinates of the radio sources from the matched coordinates of the optical counterparts. We averaged the offset to determine a systematic relative offset $\rm \Delta RA_{sys} = 0.10$\arcsec, $\rm \Delta Dec_{sys}=0.18$\arcsec\ for \diskfrb\ and $\rm \Delta RA_{sys} = 0.12$\arcsec, $\rm \Delta Dec_{sys}=0.18$\arcsec\ for \halofrb. 

The full positional error is taken as the quadrature sum of the statistical and the systematic errors. The burst position of \diskfrb is J2000 $\rm RA=00h42m01.734s$ and $\rm Dec=+41^\circ25$\arcmin$02.4143$\arcsec and $\rm \Delta RA = 0.10$\arcsec and $\rm \Delta Dec = 0.18$\arcsec. For \halofrb, the burst position is  J2000 $\rm RA=00h48m23.9608s$ and $\rm Dec=+42^\circ00$\arcmin$21.9249$\arcsec and $\rm \Delta RA = 0.12$\arcsec and $\rm \Delta Dec = 0.18$\arcsec. The localization of the bursts is shown in Figure \ref{fig:cleandirty}.

\subsection{Follow-up GBT observation}
Follow-up observations of the repeating \halofrb\ were done using the 100-m Robert C. Byrd Green Bank Telescope (GBT) on MJDs 60360, 60363, 60367, 60368, 60371, 60373, 60374, and 60422. The VEGAS pulsar mode backend recorded the data in 8-bit \textsc{PSRFITS} format. The data has a center frequency of 1.4 GHz, bandwidth of 800 MHz, time resolution of 81.92 $\mu$s and a frequency resolution of 195 kHz (4096 channels). Full polarization data were recorded in Stokes IQUV format. Bright quasars B2209+080 (only on MJD 60360) and 3C48 (on MJDs 60363 and 60371) were observed as flux calibrators and test pulsars B0531+21 ad B1933+16 (only on MJD 60363) were observed for verifying calibration. 

We cleaned the data to remove any radio frequency interference (RFI) using a custom filter that uses Savitzky-Golay and Spectral Kurtosis \citep{nita2010} filter. We searched the GBT data using \textsc{Your} \citep{Your2020} package which uses \textsc{Heimdall} \citep{barsdell2012heimdall} to perform a single pulse search. We searched the data at different trial DMs between 600 -- 900 \dmunits. The \textsc{Heimdall} candidates were then classified into real and RFI signals using the machine learning classifier \textsc{Fetch} \citep{Agarwal2020}. 

We did not detect any bursts in a total of 5.7 hours on source above $7\sigma$. From the \realfast detection, we calculate the FRB burst rate to be $\rm 0.16~hr^{-1}$ above a flux limit of 29 mJy. The non-detection with GBT is therefore consistent with the burst rate of the FRB, assuming a Poissonian distribution. 

\subsection{VLA observation of M31}
We also use the primary data output of 20A-346 from the Local Group L-band Survey (LGLBS; E. Koch et al., submitted) to check for detections of the host galaxies in the radio-continuum and 21-cm \hi\ emission, and to place limits on the \hi\ column density in these sightlines through M31.
We use 43 M31 tracks from 20A-346, all taken in the VLA's B configuration\footnote{Additional A and B configuration observations are included in LGLBS, but calibration and quality assurance for these data remain on-going.}.
Each track includes a single 4~min scan on each of the 49 pointings in the complete LGLBS M31 mosaic.
Here we only use field \texttt{M31LARGE\_14}, which is closest to the \diskfrb\ location, and \texttt{M31LARGE\_47}, which is closest to the \halofrb\ location. After accounting for lost observing time due to RFI and antenna slewing, the total on‑source integration was approximately 2.75 hr.
We combine all tracks and image the source within $1.3\mbox{--}1.5$~GHz from the L-band coverage, where there is minimal RFI at the VLA site, using \texttt{tclean} with robust-0.5 weighting and the wproject gridder.

For the \diskfrb\ (field \texttt{M31LARGE\_14}), the continuum image is dynamic range limited using standard calibration techniques due to a $0.28$~Jy/beam source that is $\approx5\arcmin$ offset from the host location. We derived phase and amplitude self-calibration solutions using the \texttt{auto-selfcal}\footnote{\url{github.com/jjtobin/auto_selfcal}} routines.
After self-calibration, the local RMS near the host location is $18$~$\mu$Jy/beam, roughly twice the theoretical noise limit of $10$~$\mu$Jy/beam.
We use the self-calibrated image deconvolved to $3\sigma\approx60$~$\mu$Jy/beam with a restored beam size of $5\arcsec\times4\arcsec$.

We detect a $250\pm71$~$\mu$Jy/beam L-band radio continuum source consistent with the host location.
We fit the radio source to a 2D Gaussian using \textsc{CARTA} and find it is moderately resolved with a size of $7.1\arcsec\times5.4\arcsec$ with uncertainties of $\sim1\arcsec$ in each direction, consistent with the pixel size of the map.

We image individual $1$MHz ($\sim50$\kms) channels after subtracting the continuum model. Using a $6\arcsec$ diameter circular aperture centered at the radio source location, we extract a spectrum from 1.32 to 1.404 GHz, corresponding to redshifts up to $z=0.08$. Lower frequencies, approaching the optical line-determined redshift of $z=0.0925$ (\S\ref{sub:host_galaxy_obs}), are severely affected by RFI. The non-detection of HI up to $z=0.08$ is therefore consistent with the host redshift of $z=0.0925$ (\S\ref{sub:host_galaxy_obs}).

Finally, we measure M31's \hi\ column density towards \diskfrb\ using LGLBS's $0.4$~\kms\ resolution coverage of the 21-cm \hi\ line from $-700\mbox{--}100$~\kms (centered at M31's systemic velocity near $-300$~\kms). We center an aperture corresponding to the synthesized beam size of $5.23\times4.81''$ and beam position angle of $-$76.8 deg on the location of the FRB in the integrated intensity image and extract a mean integrated intensity of 574 K km s$^{-1}$, corresponding to an \hi\ column density of 1.05$\times10^{21}$ cm$^{-2}$ under the optically thin assumption. 

For the \halofrb\ (field \texttt{M31LARGE\_47}), we do not detect a radio continuum source and set a $5\sigma$ upper limit of $50
~\mu$Jy/beam. The locally-measured rms of $10~\mu$Jy/beam is consistent with the expected theoretical noise, and thus we did not use self-calibration for the continuum imaging. We estimate a \hi\ column density of $\rm 1.7\times10^{21} cm^{-2}$ towards the sightline of \halofrb, measured by taking the mean pixel value in an integrated intensity image within an aperture matching the size of the synthesized beam ($5.1\arcsec\times4.8\arcsec$) that is centered on the sightline. We then integrated over velocity range of -700 km/s to +100 km/s LSR. The $1\sigma$ rms in the \hi\ column density over this velocity range, for both the FRB lines-of-sight, is equal to $\rm 2.2\times10^{20}\,cm^{-2}$.

\subsection{Host Galaxy Optical Observations} 
\label{sub:host_galaxy_obs}
To accurately determine the DM contribution from the intergalactic medium, it is necessary to obtain redshifts for the FRB host galaxies. Host galaxy observations also help us understand the progenitors of FRBs and their formation channels \citep{Bhandari2022,Gordon2023a,Law2024,Sharma2024, Shannon2024, Bhardwaj2024}. In this subsection, we describe the optical identification and follow-up of the host galaxies of \diskfrb\ and \halofrb.

\subsubsection{Host Galaxy of \diskfrb}
\begin{figure*}[ht]
\centering
\gridline{\fig{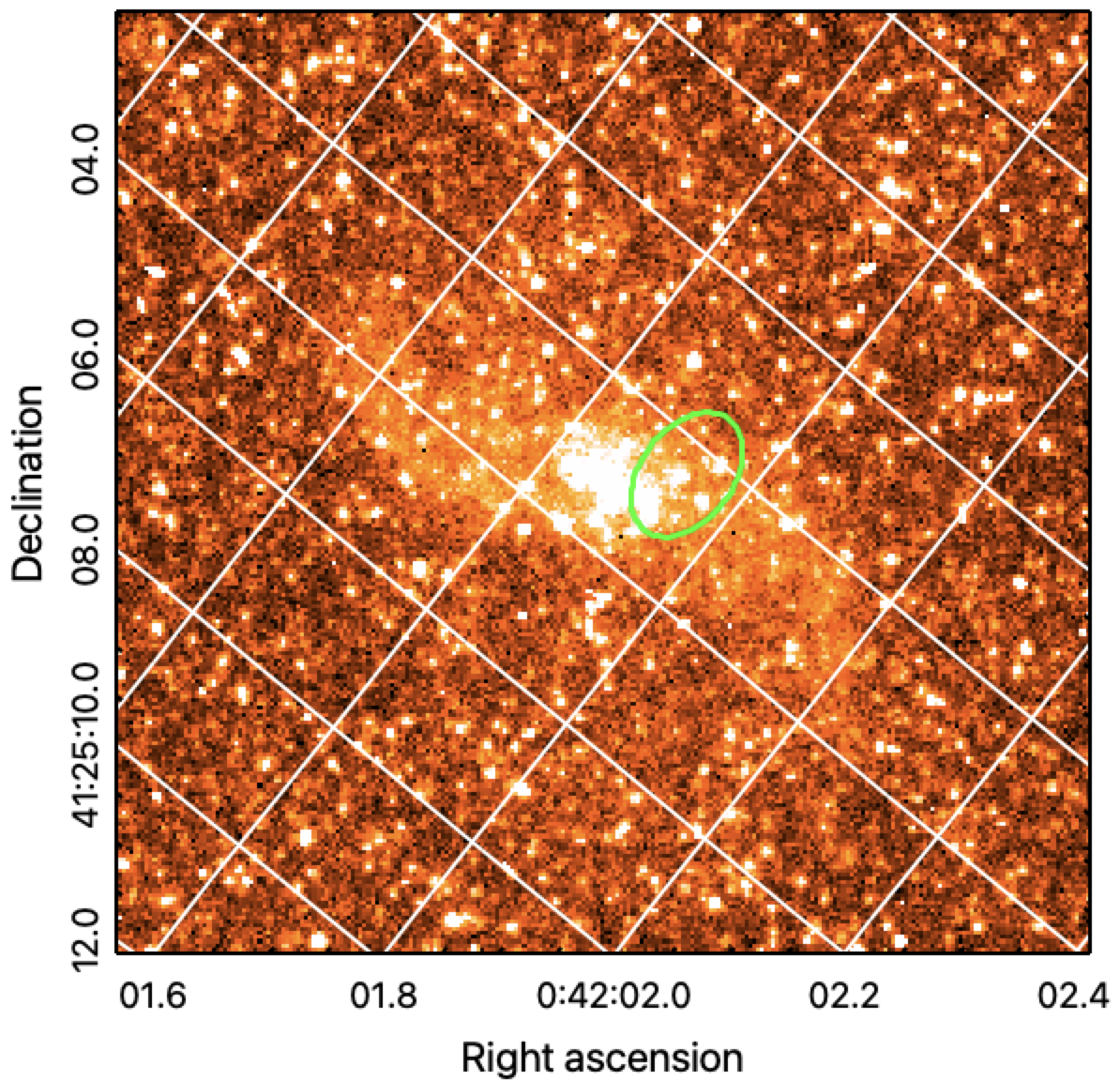}{0.4\textwidth}{(a)}
          \fig{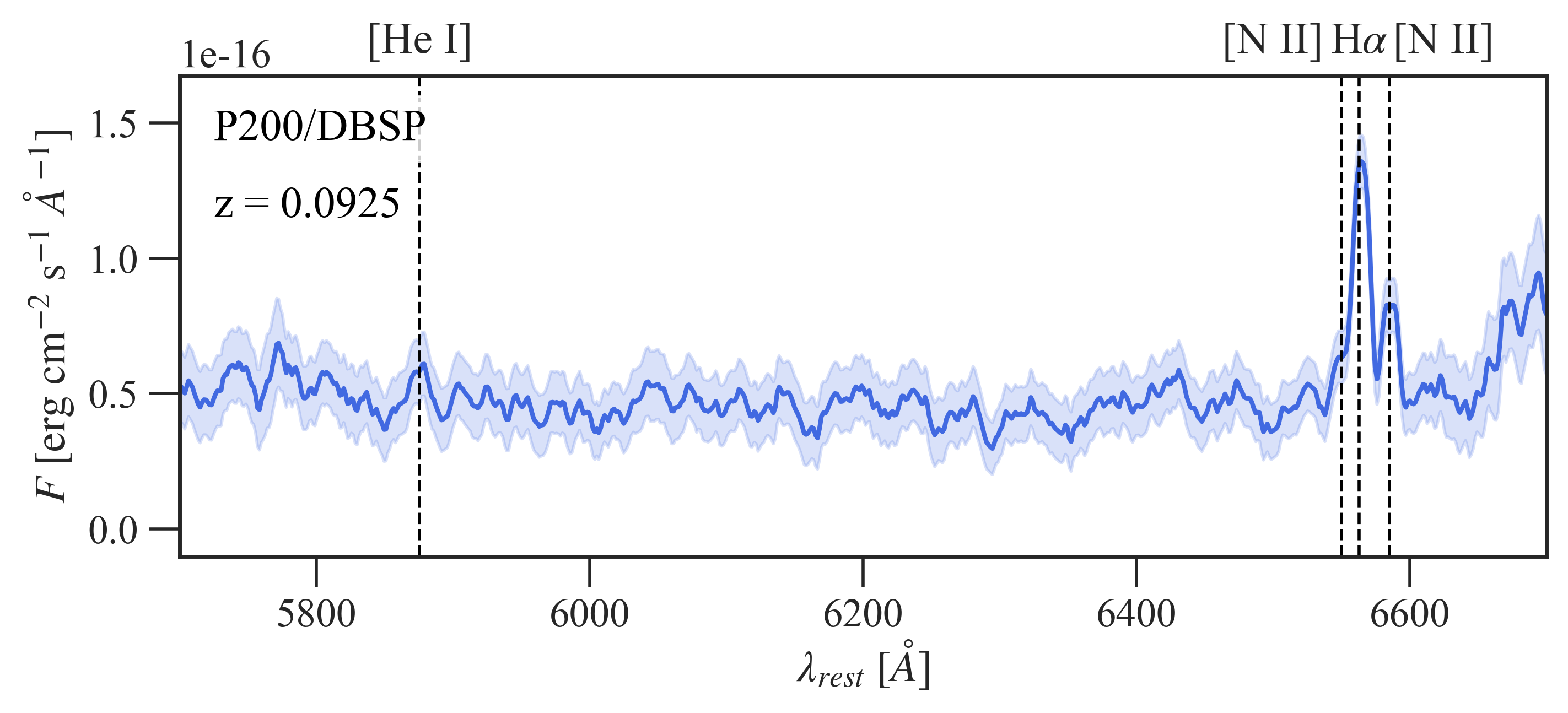}{0.6\textwidth}{(b)}
          }
\caption{\emph{(a)} The HST/WCS image of the host galaxy of \diskfrb. The 5$\sigma$ \realfast localization region is shown in green. \textit{(b)} The DBSP spectrum of the host galaxy of \diskfrb.}
\label{fig:hg}
\end{figure*}

We associate the \diskfrb\ with the galaxy coincident with the FRB location, a spiral galaxy at R.A. = 00h42m01.676s, Dec = $41^\circ25$\arcmin3.143\arcsec, cataloged as WISEA J004201.69+412502.9 in the NASA Extragalactic Database and PSO J010.5070+41.4175 in the Pan-STARRS catalog. We employed \textsc{Astropath} \citep{Aggarwal2021} to determine the association probability of the FRB and host galaxy. We find that this galaxy has a posterior of 0.9994 adopting standard priors for the offset distribution of FRBs \citep{Shannon2024}, thus confirming the association. This galaxy was also detected in the archival Hubble Space Telescope (HST) data. The HST/ACS image was CR-cleaned, Gaia-aligned, and the calibrated drizzled images in the optical photometric bands F555W and F814W is shown in Figure \ref{fig:hg}. The Milky Way Galactic extinction along this sightline is $E(B-V) = 0.086$~mag. We use the PS1 $r$-band images to subtract stellar confusion and estimate a galaxy magnitude of$r = 18.706 \pm 0.065$~mag (corrected for Galactic dust-extinction). 

Since PSO J010.5070+41.4175 lacks an archival spectrum, we followed it up with the Double Spectrograph~\citep[DBSP;][]{1982PASP...94..586O} mounted on the 200-inch Hale Telescope at the Palomar Observatory. The spectrum was obtained as a 1\arcsec\ single-slit observation on June 09, 2024 UTC with an average seeing of 1.6\arcsec\ during the observations. The 2D-spectrum shows a clear velocity gradient along the slit. The data were reduced using the DBSP\_DRP~\citep{dbsp_drp:joss} software 
\citep[built on top of the PypeIt software package;][]{pypeit}
and flux calibrated using the observations of a standard star obtained on the same night of observations. We measure the spectroscopic redshift of this galaxy using the Penalized PiXel-Fitting software~\citep[pPXF][]{2022arXiv220814974C, 2017MNRAS.466..798C}, where we jointly fit the stellar continuum and nebular emission using the MILES stellar library~\citep{2006MNRAS.371..703S}. We fit the H$\alpha$ complex to measure a redshift of $z=0.0925\pm 0.0002$. The Galactic extinction corrected H$\alpha$ flux is $1.16\times10^{-15}\rm erg\,s^{-1}cm^{-2}$.
We use \citet{osterbrock2006} to estimate a star formation rate (SFR) from the H$\alpha$ luminosity, which is shown in Table \ref{tab:hosts}.

To estimate the stellar mass, we used the PS1 \textit{r}-band image to estimate that there is 1.7 magnitudes of stellar foreground confusion toward the host galaxy. We subtract this term from the PS1 cataloged \textit{griz} photometry and use the \texttt{kcorrect} Monte Carlo sampling of the photometric errors to estimate a mean stellar mass of $9.1^{+1.0}_{-1.6}\times10^{9}~\rm M_{\odot}.$

\subsubsection{\halofrb}
\begin{figure*}[ht]
\centering
\gridline{\fig{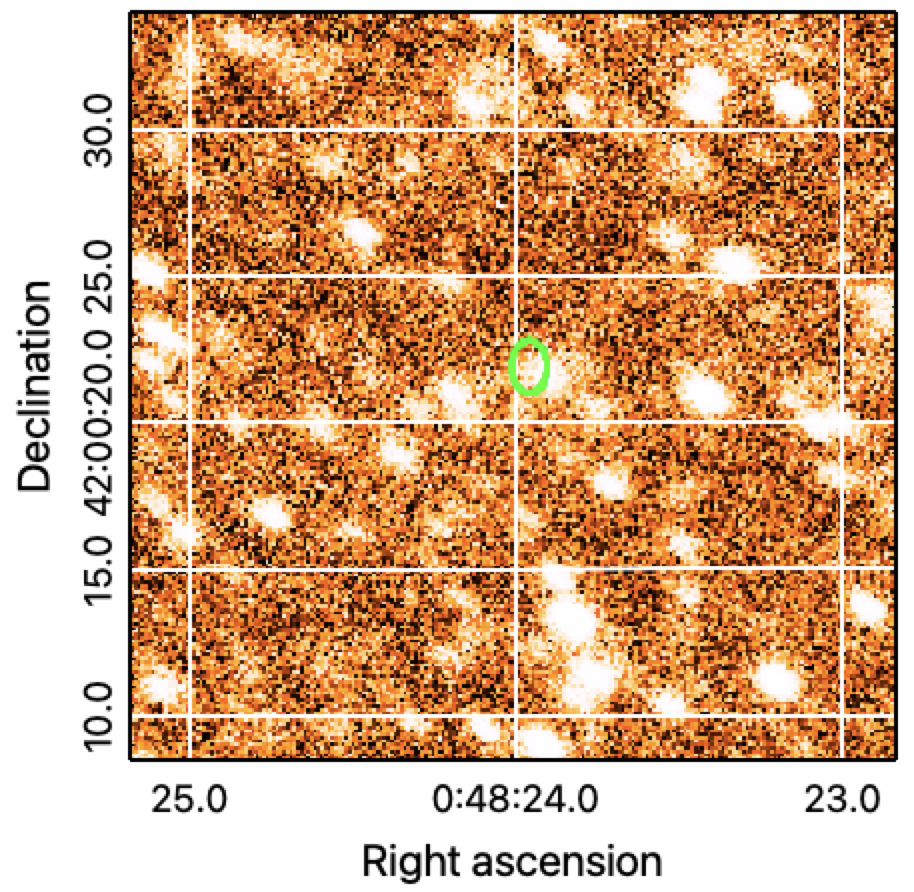}{0.4\textwidth}{(a)}
          \fig{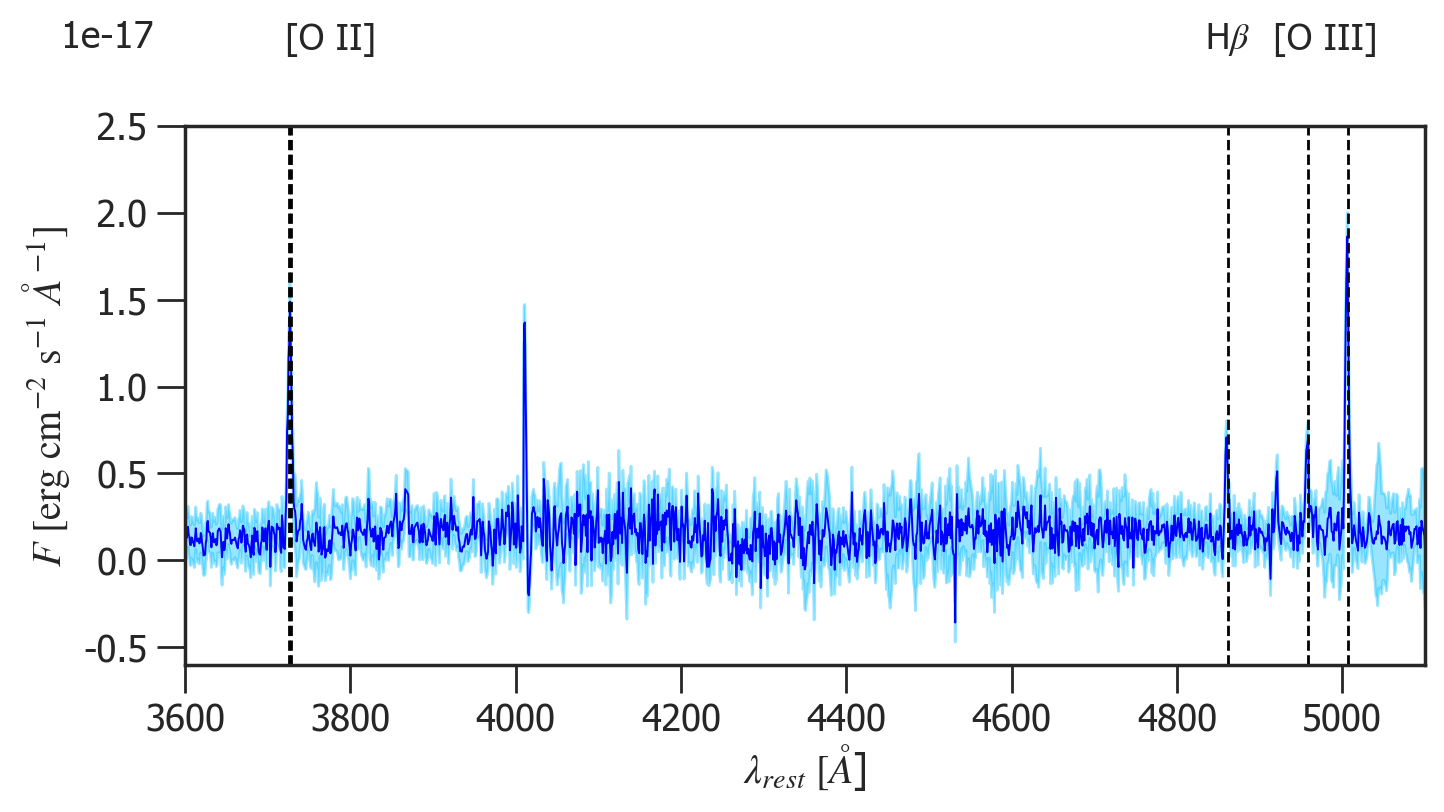}{0.6\textwidth}{(b)}}  {
          \fig{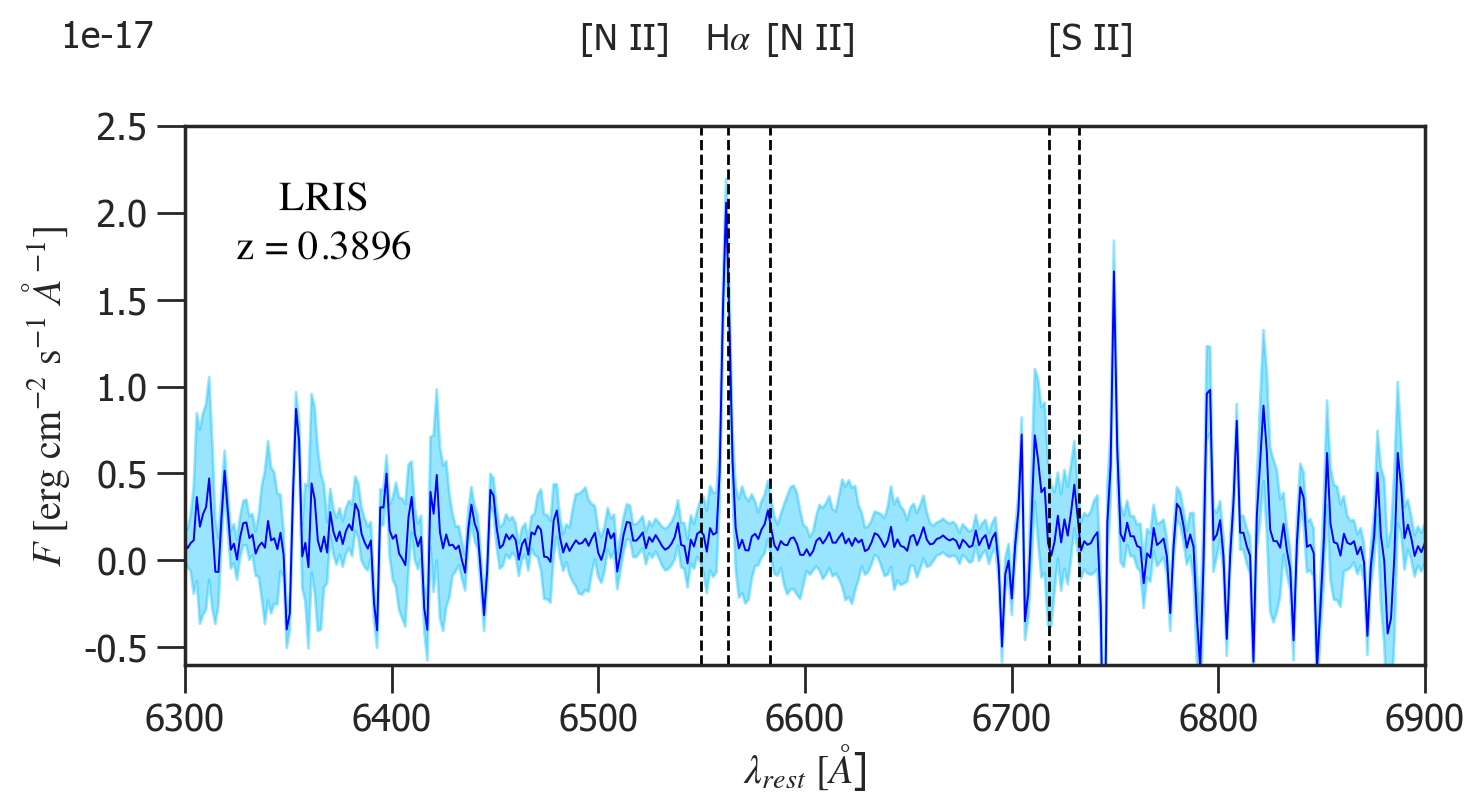}{0.6\textwidth}{(c)}}
\caption{\emph{(a)} The Keck \textit{R}-band image of the host galaxy of \halofrb. The 5$\sigma$ \realfast localization region is shown in green. \textit{(b)} The blue region and \textit{(c)} the red region, of Keck LRIS spectrum of the host galaxy of \halofrb. The redder spectrum had a poor subtraction of telluric features, but we limit our analysis to the features labeled with dashed lines.}
\label{fig:hghalo}
\end{figure*}

We obtained imaging of the field of \halofrb\ on September 5, 2023 UTC, using Deep Imaging Multi-Object Spectrograph on the 10m Keck II Telescope (DEIMOS; PI Gordon, Program O438; \citealt{DEIMOS}). The observations totaled 6x300s in $R$ band. However, due to high humidity and the crowded nature of the field, guiding proved difficult and thus only three of the images were reliable enough for reduction and analysis. The data were reduced using a custom pipeline based on \textsc{fswarp} \citep{swarp}. The final image had a limiting magnitude of 24.6. We detected two host candidates close to the localization region at R.A., Decl. = 00:48:23.9121, 42:00:21.954 and 00:48:24.2010, 42:00:21.059, respectively. Using a 1\arcsec\ aperture, we measure a magnitude for the best host candidate of R$_{\rm{AB}}$ of 22.8 mag (an estimate only, given the crowded field). We present the R-band Keck image of the field of \halofrb\ in Figure~\ref{fig:hghalo}. 

To confidently determine the host galaxy of \halofrb\ in the crowded field, we employed \textsc{astropath} on the Keck R-band image to calculate the probability of its association with the two nearest galaxies. We find the host galaxy of \halofrb, the one coincident with the burst, has a posterior of 0.9722 adopting standard priors for the offset
distribution of FRBs \citep{Shannon2024}. 
The second most likely galaxy has a negligible posterior of $1.095\times10^{-4}$. The probability that the host is undetected given the depth of the Keck image is similarly negligible. Thus, we conclude the host association to be robust.

We next obtained a spectrum with the Binospec spectrograph at the 6.5m MMT Observatory (PI Nugent, Program UAO-G200-24A; \citealt{fabricant_2019_pasp}) on June 8, 2024 UTC. The slit was placed and aligned to cover the location of both host candidates. We obtained 4x900s of exposure using a one arcsecond slit with the LP3800 filter, 270 lines/mm grating, and a central wavelength of 6500 Angstroms. Similarly to the DEIMOS imaging, the data were plagued by poor observing conditions. When combined with a small astrometric error that shifted the position of the candidates partially out of the slit, we could not detect any emission from either of the host candidates.

We conducted a second spectroscopic observation with the Low Resolution Imaging Spectrometer (LRIS; \citealt{LRIS}) on the 10m Keck I telescope on September 6, 2024 UTC (PI Ravi, Program C382). Observing conditions were excellent and seeing was near 0.5\arcsec. We obtained 2x1800s and 4x900s exposures with the blue and red detectors, respectively. The blue spectrum is produced with a grism with 400 lines/mm at 3400\AA\ and the red side by a grating with 400 lines/mm at 8500\AA. The slit was oriented to cover the first and second most probable host galaxy. Spectra were calibrated with \textsc{lpipe} \citep{Perley_2019}. The 1d spectral extraction was done with a boxcar and integrated over the slit.

Figure \ref{fig:hghalo} shows the spectrum and model fits. Both continuum and strong line emission are evident throughout the spectrum. As before, the best-fit model and redshift were found with the spectral modeling code \textsc{ppxf}. Based on fits with strong emission lines from the Balmer series, $\left[\rm{O II}\right]$, and $\left[\rm{O III}\right]$, we find a host redshift of 0.3896$\pm$0.0002. The spectrum for the second closest candidate host shows no significant spectral features.

Table \ref{tab:hosts} shows the line fluxes for the best-fit model. Only lines with a significance greater than 5 are shown. Using the ratio of H$\alpha$ to H$\beta$, we estimate an extinction \textit{E(B-V)}$=0.064$. The ppxf modeled line fluxes are corrected for extinction using \textsc{dust\_extinction} \citep{Gordon2024}.

We estimated the stellar mass using \textsc{Prospector} \citep{prospector2021} spectral energy distribution modeling, jointly fitting the observed photometry and spectroscopy and assuming a delayed $\tau$ parametric star formation history. We find the stellar mass to be $\rm \log M_{*} = 8.91^{+0.20}_{-0.19}~\rm M_{\odot}$. The SFR and sSFR for this galaxy are also listed in Table \ref{tab:hosts}.


%


\begin{table*}[]
    \centering

    \begin{tabular}{ccc}
    \hline
        Properties & HG \diskfrb & HG \halofrb\\
        \hline
        Alt.\ Name & PSO J010.5070+41.4175 & -  \\
        RA & 00h42m01.676s  & 00h48m23.9121s \\
        Dec& $41^\circ25\arcmin3.143\arcsec$&$42^\circ00\arcmin21.954\arcsec$\\
        $z$  & 0.0925 $\pm0.0002$ & 0.3896$\pm0.0002$ \\
        $\rm M_{*} (M_{\odot})$ & $9.1^{+1.0}_{-1.6}\times10^{9}$ & $(8.1\pm1.5)\times10^8$\\
        $\rm L_{H\alpha}~(erg\,s^{-1})$ & $3.4\times10^{40}$ & $7.0\times10^{40}$ \\
        SFR ($\rm M_{\odot}yr^{-1}$) & 0.19&0.38 \\
        sSFR ($\rm yr^{-1}$) & $(2.05\pm0.36)\times10^{-11}$ & $(4.7\pm0.9)\times10^{-10}$\\
        H$\alpha~(\rm erg\,cm^{-2}s^{-1})$&$1.48\times10^{-15}$ & $1.24(6)\times10^{-16}$\\
         H$\beta~(\rm ergs\,cm^{-2}s^{-1})$& - & $4.4(3)\times10^{-17}$\\
        $\left[\rm{O II}\right] 3726$ $~(\rm ergs\,cm^{-2}s^{-1})$& - & 4.5(5)$\times10^{-17}$ \\
        $\left[\rm{O II}\right] 3729$ $~(\rm ergs\,cm^{-2}s^{-1})$ & -&6.7(5)$\times10^{-17}$ \\
        $\left[\rm{O III}\right] 5007$ $~(\rm ergs\,cm^{-2}s^{-1})$ & -&1.6(6)$\times10^{-16}$ \\
        $\rm M_{HI} (M_{\odot})$$(z=0.08)$ & $<1\times10^{10}$ & - \\
        $\rm N_{HI}~ (cm^{-2})$ & $1.05\times10^{21}$& $1.7\times10^{21}$\\        
    \hline
    \end{tabular}

    \caption{Observed and derived properties of the host galaxies of the FRBs. Emission‑line fluxes for the host of FRB 20230506C have been corrected for the measured Balmer decrement and Galactic extinction; for the host of FRB 20230930A, only Galactic extinction was applied.}
    \label{tab:hosts}
\end{table*}


\section{DM budget} \label{sec:dm_budget}
The DM contribution from M31 can be estimated using the DM budget of the FRBs. The total DM measured for the FRBs can be written as:
\begin{equation}\label{dmeq}
\begin{split}
    \rm DM_{obs}=DM_{MW,disk}+DM_{MW,halo}+DM_{M31}\\+\rm DM_{IGM}+\frac{DM_{host}}{1+\textit{z}}\\
\end{split}
\end{equation}
Therefore, by modeling the DM contribution from disk and halo of Milky Way ($\rm DM_{MW, disk}$ and $\rm DM_{MW, halo}$), intergalactic medium ($\rm DM_{IGM}$) and the host galaxy ($\rm DM_{host}$), we can estimate the DM contribution from M31 ($\rm DM_{M31}$). 
The probability density function (PDF) of the sum of independent variables can be written as the convolution of the PDF of each variable. Therefore from equation \ref{dmeq}, we can write:
\begin{equation}
\label{eq:dm_pdf}
\begin{split}
         \rm \mathcal{P}(DM_{M31})=\mathcal{P}(DM_{obs})*\mathcal{P}(-DM_{MW,disk})\\ \rm *\mathcal{P}(-DM_{MW,halo})
    *\rm \mathcal{P}(-DM_{IGM})\\\rm *\mathcal{P}(-DM_{host})
\end{split}
\end{equation}

Since the FRBs have asymmetric uncertainties on their observed DM values, we model $\rm \mathcal{P}(DM_{obs})$ as a split-normal PDF where $\mu$ is the central value of the DM and $\sigma_1$ is the standard deviation below $\mu$ and $\sigma_2$ is the standard deviation above $\mu$. Here, for \diskfrb, we take $\mu=456$ \dmunits, $\sigma_1=0.6$ and $\sigma_2=0.5$ \dmunits and for \halofrb, $\mu=772$ \dmunits, $\sigma_1=2$ and $\sigma_2=3$ \dmunits.

\subsection{DM from Milky Way}
The Milky Way disk contribution to the DM in the line of sight  of the FRBs can be obtained from electron density distribution models like NE2001 \citep{ne2001}. For \diskfrb\ the $\rm DM_{MW,disk}$= 70 \dmunits\ and for \halofrb it is 68 \dmunits\ from the NE2001 \citep{ne2001} model. To constrain the uncertainty on the NE2001 model estimation, we examined the DMs of nearby pulsars. The closest pulsar to both FRBs ($\sim 6^{\circ}$ away) J0039+35 has a DM of 53 \dmunits\ and this sets a lower limit to the line of sight Milky-Way disk contribution. The distance to this pulsar remains unconstrained, rendering it ineffective for reducing the uncertainty in the NE2001 model estimation. 
To estimate the uncertainty on the Milky Way DM from the NE2001 and YMW16 \citep{ymw16} models, we then identified pulsars from the PSR$\pi$ sample \citep{Deller2019}, with independent distance measurements, that were within $\sim 20^\circ$ of the position of the FRBs. We found three pulsars J0040+5716, J0055+5117, and J0147+5922	within this region. Although NE2001 slightly underestimates the DM for J0147+5922, it over predicts the DM of the other two pulsars by a factor of two, until their measured distances (see Table \ref{tab:psrs}). It is worth noting that J0147+5922 has the lowest Galactic latitude ($b=-2.7$) of all the three pulsars and the FRBs.
\begin{table}
    \centering
    \begin{tabular}{ccccc}
    \hline
        Pulsar &$\rm DM_{measured}$ & Distance & $\rm DM_{predicted}$ & \\
         & $\rm (pc\,cm^{-3})$ & (kpc) &  $\rm (pc\,cm^{-3})$ & \\
         \hline
         J0040+5716& 92.6 & 9.77 & 183.3  \\
         J0055+5117& 44.1 & 2.87 & 71.1 \\
         J0147+5922& 40.1 & 2.02 & 31.8  \\
         \hline
    \end{tabular}
    \caption{PSR$\pi$ pulsars within $20^\circ$ of the line of sight of both FRBs. The predicted DM is from NE2001 model.}
    \label{tab:psrs}
\end{table}
 
To account for these systematic discrepancies, we assume a 30\% uncertainty on the Milky Way disk DM contribution.  Therefore, $\rm \mathcal{P}(DM_{MW,disk})$ is a truncated Gaussian distribution with a lower limit set at 53 \dmunits,  $\mu=70$ \dmunits\ and $\sigma = 21$ \dmunits\ for \diskfrb\ and $\mu=68$ \dmunits\ and $\sigma = 20$ for the \halofrb. For the Milky Way halo, we assume a Gaussian distribution with $\mu=38$  \dmunits\ and $\sigma=19$ \dmunits\ corresponding to a 50\% uncertainty \citep{Ravi2023}.

\subsection{DM from the IGM}
We calculate the average $\rm DM_{cosmic}$ using equation (2) in \cite{Macquart2020}. We use the same prescription to calculate the probability density and use the values $\alpha=3.0$ and $\beta=3.0$ for the parameters describing the inner halo density profile and feedback parameter $F=0.31$. 


\subsection{DM from host galaxies}\label{sec:dmhost} Although host galaxies are expected to contribute significantly to the total DM of FRBs, estimating this contribution is challenging. H$\alpha$ surface density can trace the emission measure and, in principle, the DM. However, this estimate depends on several factors—such as the gas temperature, the clumpiness of the H$\alpha$-emitting regions, and the FRB’s path length through the host—resulting in a potentially wide range of $\rm DM_{host,disk}$ that can overlap with the observed total DM.

Using a large sample of cosmological FRBs, \cite{Connor2024} found that $\rm DM_{host}$ is well described by a log-normal distribution with $\mu=4.90^{+0.18}_{-0.20}$ and $\sigma = 0.53^{+0.16}_{-0.14}$. We therefore adopt the same log-normal prior for our host galaxies. This also includes the contribution from the host's halo.
\begin{figure*}
    \centering
    \includegraphics[width=1\linewidth]{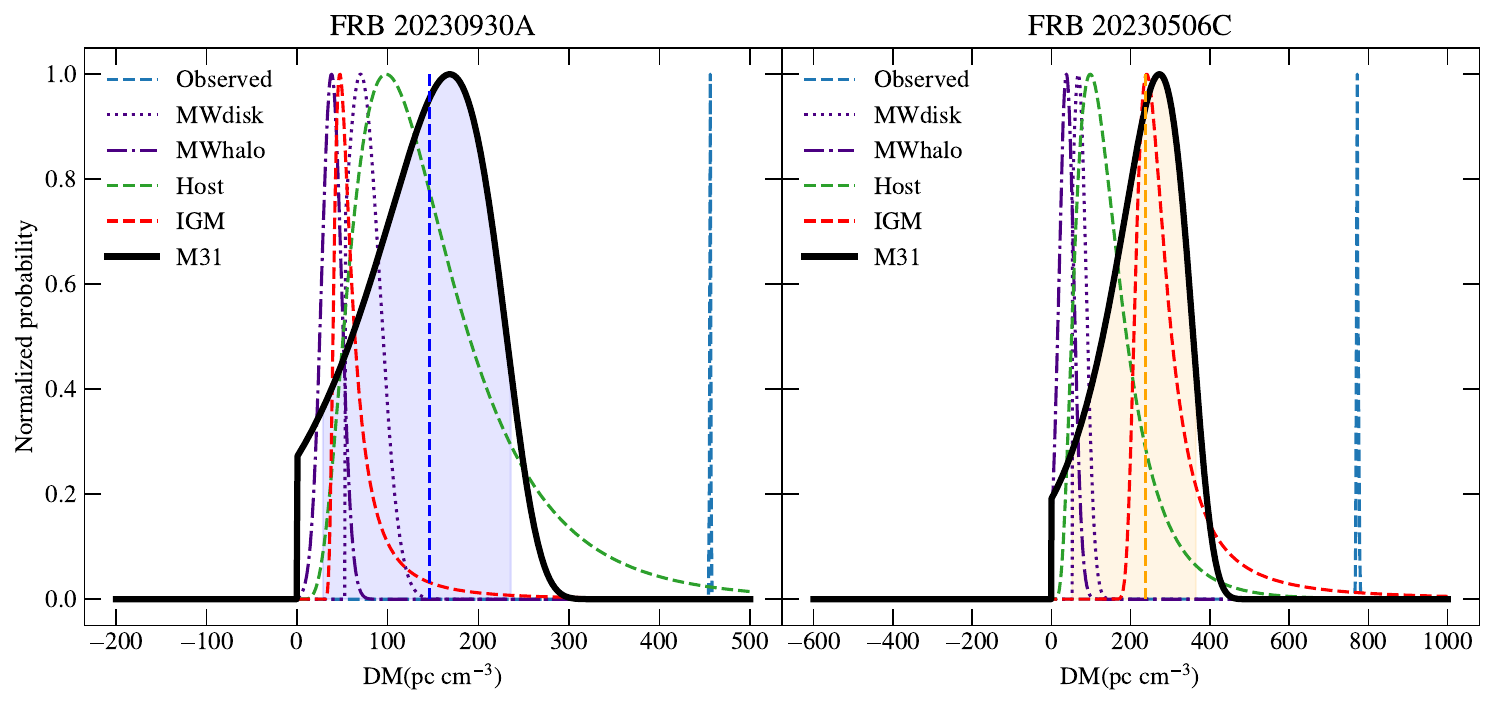}
    \caption{The probability distribution of all different DM components. The \emph{left} panel corresponds to \diskfrb\ and the \emph{right} panel to \halofrb. The blue line shows the PDF of the observed DM. The indigo dotted line represents the MW disk contribution, while the dash–dot line represents the MW halo. The green dashed line corresponds to the host DM distribution, the red line to the IGM distribution, and the black line to M31. In the left panel, the dark blue dashed vertical line marks the median of the $\rm DM_{M31}$ distribution, with the shaded region indicating the 90\% CI. In the right panel, the orange dashed vertical line marks the median, and the shaded region likewise denotes the 90\% CI}.
    \label{fig:dm_pdf}
\end{figure*}

\subsection{DM from M31}
To convolve different PDFs, it is essential to assume that each PDF is independent of the others. To verify this assumption, we performed Monte Carlo sampling for each DM component from its respective distribution and computed the pairwise Pearson correlation coefficients. The largest correlation coefficient was found to be $0.003$, indicating that the DM components can be considered independent.

We defined the PDF of each component as described above, imposing a zero probability for negative DM values, followed by normalization. 
The PDF of $\rm DM_{M31}$ was then computed via the convolution in Equation \ref{eq:dm_pdf}.  probability of any negatives in $\rm DM_{M31}$ were again set to zero and the distribution renormalized, from which the median and CIs were derived. We adopt a 90\% CI, rather than the more commonly used 68\%, to provide a more conservative characterization of our asymmetric posterior distribution. The resultant distribution is shown in Figure \ref{fig:dm_pdf} and the values are listed in Table \ref{tab:dm_combined_results}.

\subsection{Independent DM constraints on M31}
\begin{figure}
    \centering
    \includegraphics[width=0.5\textwidth]{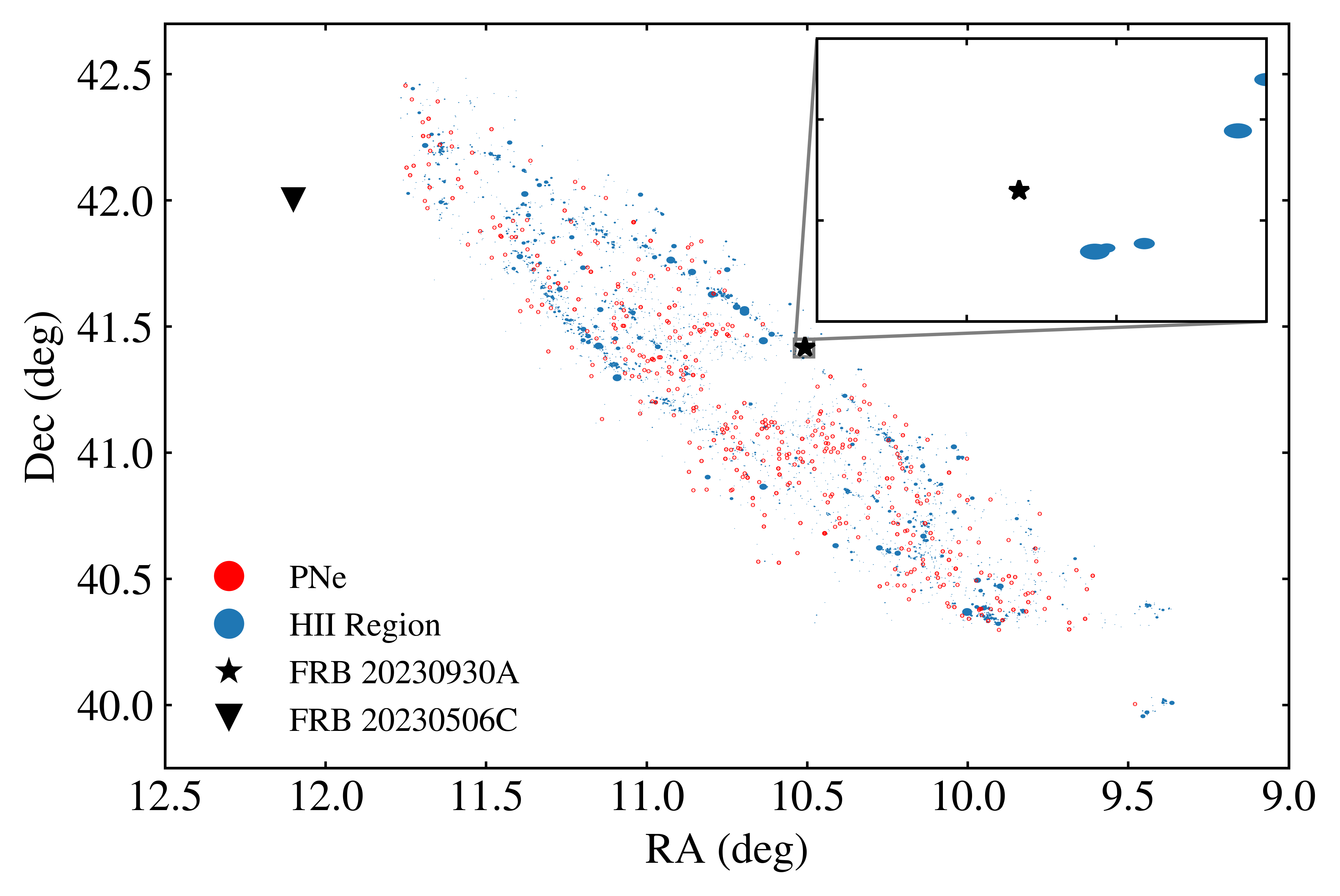}
    \caption{The map of HII regions and Planetary nebula of M31 with respect to the FRB lines of sights. The region zoomed around \diskfrb\ is shown in the inset. The size of the HII regions and planetary nebulae circles are scaled by their actual size. The catalog is adapted from \cite{Azimlu2011}.}
    \label{fig:h2reg}
\end{figure}

 The M31 DM we determined in the above section is the sum total of the DM contribution from its disk and halo. In this section, we identify the individual contribution to the total DM from the disk and halo. To make sure that the FRB line of sight is not intersecting any regions of excess electron density, we compared the positions of the FRBs with respect to the HII regions of M31. \cite{Ocker2024} has revealed that HII regions can contribute tens to hundreds of DM units depending up on the path length intersecting the region. It is clear from the Figure \ref{fig:h2reg}, that the FRB sight-lines do not intersect any cataloged HII regions or planetary nebulae in M31 \citep{Azimlu2011} and therefore the $\rm DM_{M31}$ might not be dominated by any over dense regions.
\subsubsection{M31 Disk DM}\label{sec:m31disk}
\begin{figure}
    \centering
    \includegraphics[width=1\linewidth]{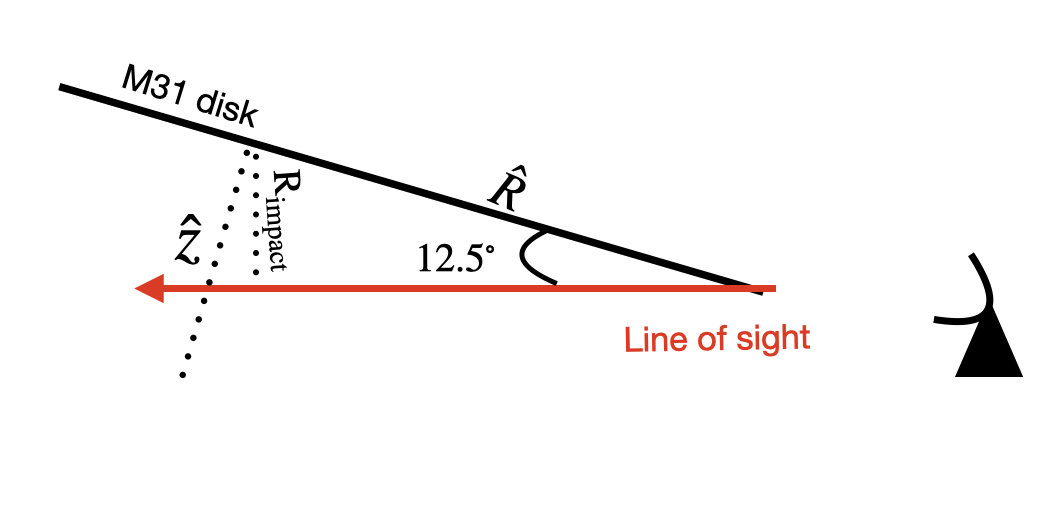}
    \caption{The geometry of M31's disk with respect to the \diskfrb's line-of-sight. The DM from M31's disk is calculated by integrating the electron density $n_e(R,z)$ along this line-of-sight.}
    \label{fig:los_geometry}
\end{figure}
We modeled the electron density distribution of the thin and thick disk of M31 as a function of radial and vertical distance, assuming it is similar to the Milky Way using the models from \cite{ymw16}. The thick disk was modeled using the equation
\begin{equation}
    n_{\rm thick} = n_{\rm thick,0} g_d {\rm sech^2}\left(\frac{z}{H_1}\right)
\end{equation}
where $n_{\rm thick,0}$ and $H_1$ is the mid plane density and scale height respectively. The vertical and the radial extent of the disk is set by the parameters $A_d$ and $B_d$. For $R \le B_d, g_d=1$ and for $R > B_d, g_d= {\rm sech^2}\left(\frac{R-B_d}{A_d}\right)$. The electron density of the thin disk is modeled by the equation 
\begin{equation}
    n_{\rm thin} =n_{\rm thin,0} g_d~ {\rm sech}^2 \left(\frac{R-B_2}{A_2}\right){\rm sech^2}\left(\frac{z}{K_2 H}\right)
\end{equation}
where $H$ is the parameterized scale height on $R$ given by:
\begin{equation}
    H = 32 + 1.3\times10^{-3}R+4.0\times10^{-7}R
\end{equation}
The Table \ref{tab:disk_model} lists the values of the constants used. 

\begin{table}
    \centering
    \begin{tabular}{ccc}
    \hline
        Constant & value & units\\
        \hline
         $A_d$& 2500 & pc \\
         $B_d$&15000 & pc \\
         $K_2$& 1.54 & \\
         $A_2$& 1200 & pc\\
         $B_2$& 4000 & pc \\
         $H_1$& 1673 & pc \\
         $n_{\rm thick,0}$& 0.01131 & $\rm cm^{-3}$\\
         $n_{\rm thin,0}$ & 0.404 & $\rm cm^{-3}$\\
        \hline
    \end{tabular}
    \caption{The assumed values for the disk modeling of M31. The values are taken from \citep{ymw16}}
    \label{tab:disk_model}
\end{table}
 The total electron density of the M31 disk is, $n_{\rm total} = n_{\rm thick} + n_{\rm thin}$. We take the center of M31 as the origin and the LOS of the FRBs are inclined to the plane of M31 at an angle of $12.5^{\circ}$ \citep{Simien1978}. For \diskfrb, a projected impact parameter of 2.7 kpc corresponds to a galactocentric radius of 12.5 kpc within M31’s disk. By integrating the total electron density $n_{\rm total}$ along the line of sight that crosses the disk vertically at this radius in the $z–R$ (Fig.~\ref{fig:los_geometry}), we estimate a disk contribution $\rm DM_{M31,disk}=128$ \dmunits. At an impact parameter of 17.1 kpc, the M31 disk contributes negligibly to the DM of \halofrb. In this modeling, we assumed that M31 is analogous to the Milky Way, and disregarded the contributions from spiral arms and other galactic components. To account for this in addition to the inherent uncertainties in the YMW16 model, we assume a 50\% uncertainty on the calculation.
\begin{figure}
    \centering
    \includegraphics[width=1\linewidth]{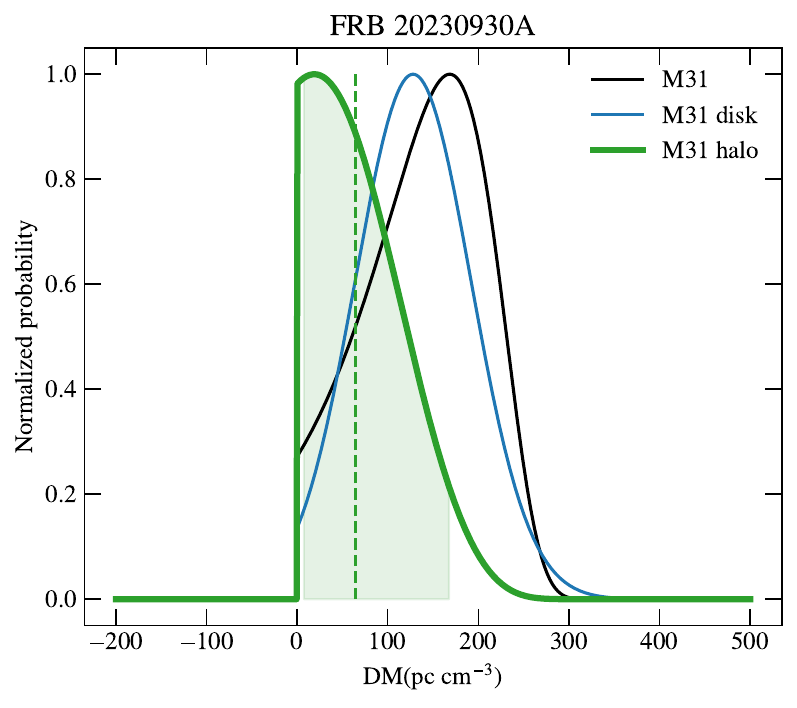}
    \caption{The $\rm DM_{M31,halo}$ estimated from \diskfrb. The total $\rm DM_{M31}$ is shown in black, with the $\rm DM_{M31,disk}$ in blue and the $\rm DM_{M31,halo}$ in green. The dashed green line marks the median of the $\rm DM_{M31,halo}$ PDF for \diskfrb, and the shaded region indicates the 90\% CI.}
    \label{fig:dmhalo}
\end{figure}

\begin{figure}
    \centering
    \includegraphics[width=1\linewidth]{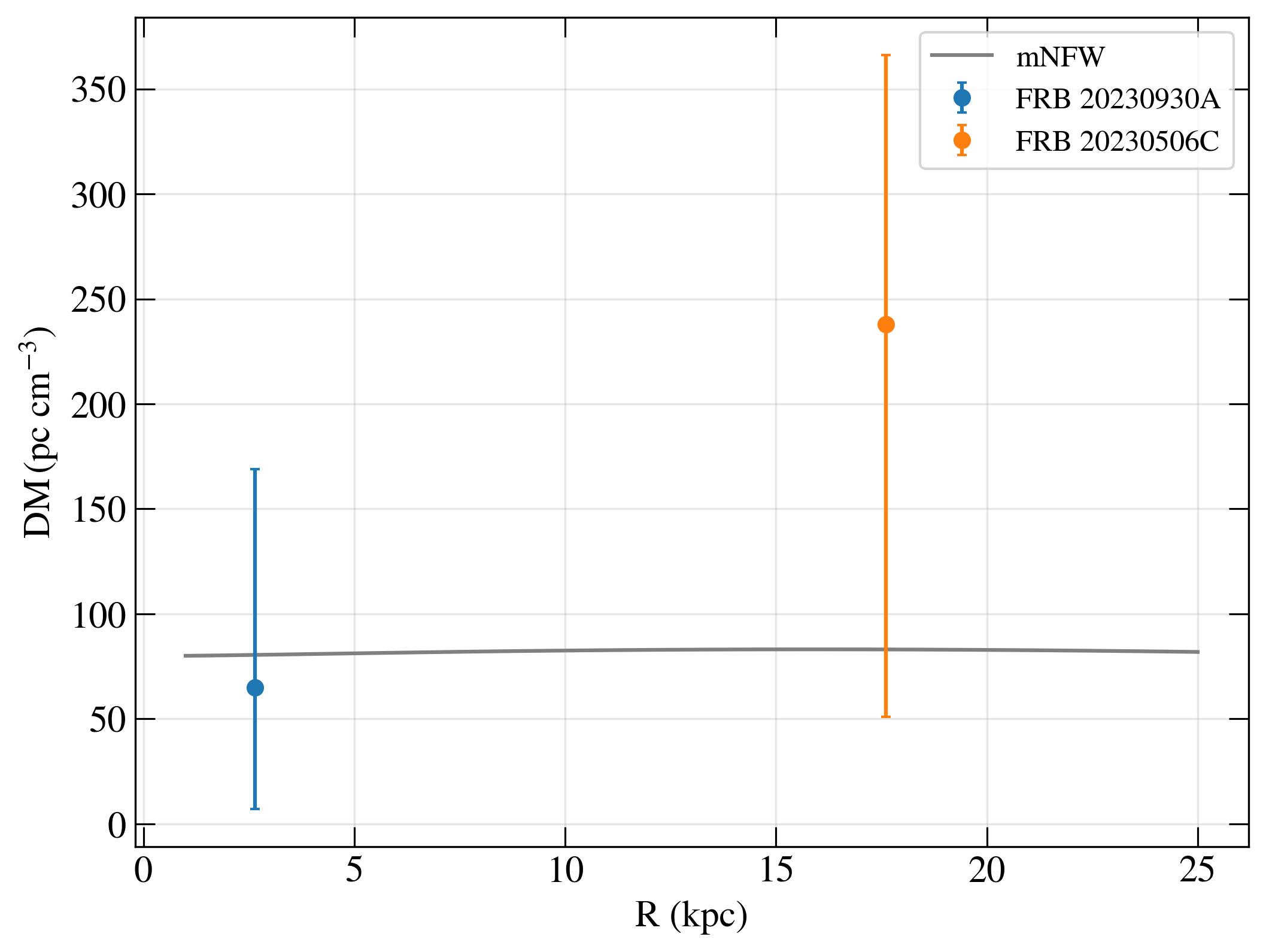}
    \caption{The DM predicted from mNFW profile of M31's halo for a given impact parameter from the center of the halo \textit{(grey)}. The $\rm DM_{M31,halo}$ measured from the two FRBs are also shown. The FRB measurements are centered on the median and extends to the 90\% CIs.}
    \label{fig:dmvsr}
\end{figure}
\subsection{M31 halo DM}\label{sec:dm_halo}

The estimation of $\rm DM_{M31,disk}$ allows us to isolate the $\rm DM_{M31,halo}$ from the total $\rm DM_{M31}$. The PDF of $\rm DM_{M31,halo}$ (see Figure \ref{fig:dmhalo}) can be written as:
\begin{equation}
    \rm \mathcal{P}(DM_{M31,halo})= \mathcal{P}(DM_{M31})*\mathcal{P}(-DM_{M31,disk})
\end{equation}
M31's contribution to the DM of the FRBs from its CGM is given in Table \ref{tab:dm_combined_results}. 

\begin{table*}
    \centering
    \begin{tabular}{cccccc}
        \hline
        FRB & \multicolumn{2}{c}{$\rm DM_{M31}$} & \multicolumn{2}{c}{$\rm DM_{M31,halo}$} \\
        \cline{2-3} \cline{4-5}
        & Value & 90\% CI & Value & 90\% CI \\
        & (\dmunits) & (\dmunits) & (\dmunits) & (\dmunits) \\
        \hline
        \diskfrb & 144 & [26, 239] & 65 & [7, 169] \\
        \halofrb & 238 & [51, 366] & -- & -- \\
        \hline
    \end{tabular}
    \caption{Comparison of $\rm DM_{M31}$ (left) and $\rm DM_{M31,halo}$ (right) as measured from two \realfast FRBs. For \diskfrb, the M31 disk contribution has been modeled and removed when computing $\rm DM_{M31,halo}$.}
    \label{tab:dm_combined_results}
\end{table*}

We then compared this estimate with the theoretical predictions of M31's halo profile from the mNFW profile \citep{Mathews2017}. From the mNFW profile, we can estimate the halo DM for the FRBs given their impact parameter \citep{prochaska2019}. The mNFW density profile is given by:
\begin{equation}
\mathrm{\rho_{b} = \frac{\rho^{0}_{b}}{y^{1 - \alpha}(y_{0} + y)^{2 + \alpha}}}
    \label{equation:7}
\end{equation}
where $y\equiv c(r/r_{200})$, where $c$ is the concentration parameter and $r_{200}$ is the virial radius, $\rho_b^0=(200\rho_c/3)c^3/f(c)$, $f(y)=ln(1+y)-y/(1+y)$, $\rho_c=9.2\times10^{-30}\rm ~g~cm^{-3}$ is the critical density for the Hubble constant $H_{0}=70~\rm km~s^{-1}$, $y_{0}$ is a feedback-dependent parameter, and $\alpha$ is a constant \citep{Mathews2017}. Following \citet{prochaska2019}, we set $\alpha = 2$ and $y_{0} = 2$, and a constant $c=7.67$ for M31. The DMs of FRBs traveling through a halo are dependent upon the impact parameter as is given by 
\begin{equation}
\mathrm{DM(R_{\perp}) = 2 \int_{0}^{\sqrt{r^{2}_{max} - R^{2}_{\perp}}} n_{e} ds}\label{equation:8}
\end{equation}.

Figure \ref{fig:dmhalo} shows the measured DM contribution from the halo of M31 using \diskfrb and Figure \ref{fig:dmvsr} shows the DM profile as predicted by mNFW. The mNFW halo DM contribution is consistent with the observed $\rm DM_{halo}$ within the CIs for \diskfrb\ and \halofrb.

\section{Discussion} \label{sec:discussion}

We have estimated the DM contribution of M31 along the line of sights probed by \diskfrb\ and \halofrb. After modeling the disk component of M31 along the line of sight of \diskfrb, we find that the halo of M31 alone contributes between 7 -- 169 \dmunits. For the \halofrb, we do not expect any contribution from the M31 disk and hence the DM excess corresponds to 51 -- 366 \dmunits. While the lower end of range is consistent with expectation from mNFW and overlaps with the distribution of $\rm DM_{M31,halo}$ from \diskfrb, the overall intervals show noticeable differences. We note that \halofrb\ is a repeating FRB hosted by a low-mass, star-forming galaxy, which may contribute a substantial $\rm DM_{host}$ \citep{Leung2025}. Additional possibilities for foreground contributions are discussed in \S\ref{sec:foreground}. 

The halo of M31 has been observed in temperatures up to $T\sim10^{5.5}~$K associated with warm ions like OVI. However, the hotter $T>10^6~$K ions are expected to be the dominant contributor to the DM of the halos \citep{prochaska2019}, which are yet to be observationally detected in M31. We looked at the possible contribution to the $\rm DM_{M31,halo}$ from cool ions. The cool phase gas in the halo can be probed using HI high velocity clouds (HVC) or cool ions like SiII, SiIII. These can trace the gas at $T\sim10^4~$K. 
The HVCs of M31 were obtained from \cite{Westmeier2008}. Given the hydrogen column density of HVCs, the electron column density can be estimated as:
\begin{equation}
   \rm  N_{e,cool} = \mu_eN_{HI,HVC}\left(\frac{1-\chi_{HI}}{\chi_{HI}}\right)~cm^{-2}
\end{equation}
where $\mu_e=1.167$ is the reduced mass for fully ionized hydrogen and helium and the hydrogen neutral fraction $\rm \chi_{HI} = M_{HI}/(M_{HI} + M_{HII}) =0.3$, which is commonly used in Milky Way halo estimates \citep{prochaska2019}. The closest HVC is separated by $0.4^\circ$ to \diskfrb\ and $0.5^\circ$ to \halofrb, yielding $\rm DM_{cool}=4$ \dmunits\ and 2 \dmunits\ respectively. If, instead, we assume a lower neutral fraction, $\rm \chi_{HI} =0.1$ \citep{Thilker2004}, the corresponding contributions increase to $\rm DM_{cool}=17$ \dmunits\ and 10 \dmunits\ for \diskfrb and \halofrb respectively. These values overlap with the lower end of the 90\% CI of the $\rm DM_{M31,halo}$ measurement from \diskfrb. We emphasize, however, that $\rm \chi_{HI}=0.1$ is only an estimate and not a direct measurement \citep{Thilker2004} and thus associated DM values should be treated as an upper limit. Consequently \cite{Lehner2015} reports that the ionization increases with the radius, making the outer halo much more ionized than the inner halo. Therefore, adopting $\rm \chi_{HI} = 0.3$ is likely more appropriate for our $\rm DM_{halo}$ estimates in the inner halo.

\cite{prochaska2019} also finds a direct correlation between the DM values from $\rm N_{HI}$ and Si ions. Therefore, we also estimate the $\rm DM_{cool}$ from the average contribution of Si ions. We have $\rm \langle N_{Si}\rangle = 2.5\times 10^{13}~cm^{-2}$ at $R < 25$ kpc \citep{Lehner2015}. Here $\rm \langle N_{Si}\rangle$ is the average of $\rm N_{SiII} + N_{SiIII}$. Si also has a covering fraction of about unity within $R < 0.2R_{\rm vir}$ \citep{Lehner2020}. To estimate the DM, we can write
\begin{align}
    \rm N_e \approx 1.2 N_{H} \\
   \rm  N_H = \frac{N(Si)}{Z\times (Si/H)_\odot}
\end{align}
The metallicity in the CGM of M31 is undetermined, however a lower limit of $Z>0.2Z_\odot$ is estimated \citep{Lehner2020} and we assume $\rm (Si/H)_\odot = 10^{-4.49}$ \citep{Asplund2009}. We get $\rm N_H = 3.8\times10^{18}~cm^{-2}$ and converting this to DM, we get $\rm DM_{cool,Si} < 2$ \dmunits, thereby confirming our conclusion that cool halo gas does not contribute significantly to the total DM. 

We also estimated the DM contribution from the warm component of M31’s CGM traced by OVI and CIV. \citet{Lehner2013} studied the cool CGM of nearby galaxies and and found that metallicities exhibit a bimodal distribution ranging from  $Z/Z_{\odot} =0.01 - 3$, with the higher end associated with massive outflows. However, studies of M31's CGM by \citet{Lehner2020} indicate that, due to the lack of nucleosynthetic effects on the abundances of elements such as Fe and C relative to Si, the metallicity of M31’s CGM is likely sub-solar but not significanlty below $1/3\,Z_{\odot}$. Additionally, ionization fractions can also vary from sightline-to-sightline; for example, OVI can originate in non-equilibrium cooling layers, turbulent mixing layers, or photoionized streams, resulting in ion fractions that can vary by factors of a few \citep{Oppenheimer2013}. Therefore, we assumed a range of values for $Z/Z_{\odot} =0.1 - 0.5$ and ionization fraction  $\chi_{\rm ion}=0.1 -0.9$. For CIV, we used the average column density within $R<25$ kpc from \citet{Lehner2015} and the solar carbon abundance $\rm (C/H)_\odot = 10^{-3.57}$ \citep{Asplund2009} and a covering fraction of 0.8 \citep{Lehner2015}, obtaining a 90\% CI upper bound of $\rm DM_{warm,CIV}<6$ \dmunits. Similarly for OVI, the solar oxygen abundance $\rm (O/H)_\odot = 10^{-3.31}$ \citep{Asplund2009} and a covering fraction of 1 \citep{Lehner2015}, we find the upper bound to be $\rm DM_{warm,OVI}<5$ \dmunits.  We note here that for OVI we assumed a covering fraction of 1 based on only two quasar sightlines from \cite{Lehner2015} and therefore the value for OVI should be considered as an upper limit.

Taken together, the cool and warm CGM components contribute at most $\sim 15$ \dmunits\ along \diskfrb\ and $\sim13$ \dmunits\ along \halofrb, for $\rm \chi_{HI}=0.3$\footnote{For $\rm \chi_{HI}=0.1$, this becomes at most $28$ and $21$ \dmunits for \diskfrb and \halofrb, respectively}. For \diskfrb, this contribution only grazes the lower edge of the 90\% CI, while for \halofrb\ it falls far short of the observed excess. Thus, the cool and warm phase gas is unlikely to account for the measured $\rm DM_{M31,halo}$. An additional component, most plausibly a hot, ionized halo, is likely required. Our results therefore provide indirect evidence for the presence of a hot gaseous halo around M31, along at least two sightlines.

\subsection{Other foreground contributions}\label{sec:foreground}
The other possibility for the estimated $\rm DM_{M31,halo}$ is the hot bridge connecting M31 and MW. 
There are detections of a large scale ($r\sim20^\circ$) X-ray and SZ bright hot plasma bridge between the MW and M31, with length 400 kpc and radius of 120 kpc \citep{Qu2021}. Even though, it is not a part of the M31 halo, the bridge could potentially contribute to the DM of the FRB. This plasma bridge has an electron number density of $\rm 2\times10^{-4} - 10^{-3} cm^{-3}$, for a length of 400 kpc, this can contribute DM = $80 - 400$ \dmunits. We note, however, that the reported X-ray and SZ detections are at $<5\sigma$ significance \citep{Qu2021}, and the geometry of the intersection strongly affects the actual DM contribution. Although the distribution of the $\rm DM_{M31,halo}$ measurement from the FRBs are broadly consistent with the theoretical predictions of mNFW, \halofrb\ exhibits only weak consistency. Therefore, it is plausible that the line of sight of \halofrb\ has intersected the plasma bridge.

The Triangulum galaxy, M33, is the third massive spiral galaxy in the local group, along with Milky Way and M31. We investigated whether M33 would contribute to the DM of these FRBs. We find that the FRBs in our sample are located approximately 14.5 degrees away from the position of M33, corresponding to a projected distance of 213 kpc. Given that the virial radius of M33’s halo is roughly 168 kpc \citep{Zacharie2017}, the FRB sightline lies well from outside the central halo region, and we therefore expect negligible contribution from M33 alone to the foreground DM, although we cannot rule out a contribution from shared plasma of M33 and M31 along the line of sight.

The Milky Way, M31, M33, and their numerous satellite galaxies, including the Magellanic Clouds, are all part of the Local Group. Theoretically, these galaxies are expected to be embedded in an ionized intragroup medium; however, no cool or hot intragroup medium has yet been detected observationally. We expect a potential degeneracy between the DM attributed to M31 and that from the intra-group medium. Nevertheless, \citet{Huang2025}, using the \textsc{HESTIA} simulations, modeled the DM from the Milky Way halo and the Local Group medium, and found only a negligible contribution (2 \dmunits) from the intra-group medium beyond that associated with the spiral galaxies and their satellites.

We note that we have not systematically searched for other foreground halos, which could contribute additional DM components. This is especially important for the \halofrb, given its larger distance.


\section{Conclusion} \label{sec:conclusion}
In this paper, we report on the \realfast discovery of two FRBs that pierce both the halo and the disk of M31. We detected repeat bursts from \halofrb. Optical follow‑up identified the host of \diskfrb\ at a redshift, $z=0.0925$ and the host of \halofrb\ at a redshift of $z=0.3896$. We used the DM budget of these FRBs to constrain the electron density distribution of M31's halo.
\begin{itemize}
    \item The Milky Way disk contribution was estimated from NE2001 model and a lower limit on the error bar was chosen based on the DM of the nearest pulsar. 
    \item After modeling out the DM contributions from the Milky Way, the IGM and the host galaxies, we isolated the total $\rm DM_{M31}$ in the lines-of-sight of both FRBs. The 90\% CIs for the total $\rm DM_{M31}$ can range from 26 -- 239 \dmunits\ and 51 -- 366 \dmunits\ for \diskfrb\ and \halofrb, respectively.
    \item We then modeled the disk of M31 using a Milky Way analog of YMW16 electron density model. After subtracting the disk contribution from the total $\rm DM_{M31}$, we obtained the PDF of the DM contribution from the CGM of M31, $\rm DM_{M31,halo}$. While the M31's disk contribution to the total DM of \halofrb\ was found to be zero, the non zero contribution to the \diskfrb constrained the $\rm DM_{M31,halo}$ along that line of sight to be between 7 and 169 \dmunits. 
    \item We compared our measurements to the predictions from the mNFW profile of M31's halo and find that it is broadly consistent within the CIs for both FRBs. 
    \item The measured $\rm DM_{M31,halo}$, along these two sightlines, is consistent with the presence of a hot gaseous halo around M31, as contributions from the cool and warm phases alone are unlikely to explain the observed values.
    \item The other possibility that can account for the excess in $\rm DM_{M31,halo}$ along \halofrb\ is if its line of sight intersects the plasma bridge between MW and M31. We also discuss the possibility of additional DM contribution from M33 and the local group medium and finds it to be negligible.
\end{itemize}

In this work, we have demonstrated how FRBs can be used to study the CGM of nearby galaxies. Higher time resolution and polarimetric follow-up of the repeating \halofrb\ will help constrain the turbulence and magnetization of the halo by measuring properties such as scattering and rotation measure \citep{Prochaska2019b}. 

Instruments such as CHIME are expected to detect and localize hundreds of FRBs intersecting M31, enabling a detailed reconstruction of its DM profile, particularly in the outer halo where our current constraints are limited. Improved estimates of $\rm DM_{host}$, are also crucial for establishing more informative priors. Integral Field Spectroscopy of FRB host galaxies \citep{Bernales-Cortes2025} offers a promising path forward, as spatially resolved maps of ionized gas can reveal whether an FRB resides within a dense, ionized clump that contributes significantly to $\rm DM_{host}$ or instead lies in a comparatively diffuse environment.

\section{Acknowledgments}
We thank the anonymous referee for valuable suggestions.
We also thank Nicolas Lehner and Zhijie Qu for useful discussions. RAT and SBS were supported in this work by NSF award \#1714897. RAT acknowledges funding for this work from the European Research Council (ERC) under the European Union's Horizon 2020 research and innovation programme (`EuroFlash'; Grant agreement No. 101098079). SBS gratefully acknowledges the support of a Sloan Fellowship. CJL acknowledges support from the National Science Foundation under Grant No.\ 2022546. \realfast\, is supported by the NSF Advanced Technology and Instrumentation program under award 1611606. Computational resources were provided by the WVU Research Computing Dolly Sods HPC cluster, which is funded in part by NSF MRI Grant \#2117575. A.C.G. and the Fong Group at Northwestern acknowledges support by the National Science Foundation under grant Nos. AST-1909358, AST-2308182 and CAREER grant No. AST-2047919. A.C.G. acknowledges support from NSF grants AST-1911140, AST-1910471 and AST-2206490 as a member of the Fast and Fortunate for FRB Follow-up team. LC is grateful for support from NSF grants AST-2205628 and AST-2107070.

The Green Bank and National Radio Astronomy Observatories are facilities of the National Science Foundation operated under cooperative agreements by Associated Universities, Inc. We thank the telescope operators and project friends during all our VLA and GBT observations.

The Local Group L-Band survey (LGLBS) is an Extra Large program conducted on Jansky Very Large Array, which is operated by the National Radio Astronomy Observatory (NRAO) and includes observations from VLA projects 20A-346, 13A-213, 14A-235, 14B-088, 14B-212, 15A-175, 17B-162, and GBT projects 09A-017, 13A-420, 13A-430, 13B-169, 14A-367, 16A-413.
Execution of the LGLBS survey science was supported by NSF Award 2205628.

W. M. Keck Observatory and MMT Observatory access was supported by Northwestern University and the Center for Interdisciplinary Exploration and Research in Astrophysics (CIERA).

Some of the data presented herein were obtained at the W. M. Keck Observatory, which is operated as a scientific partnership among the California Institute of Technology, the University of California and the National Aeronautics and Space Administration. The Observatory was made possible by the generous financial support of the W. M. Keck Foundation.

The authors wish to recognize and acknowledge the very significant cultural role and reverence that the summit of Maunakea has always had within the indigenous Hawaiian community.  We are most fortunate to have the opportunity to conduct observations from this mountain.

Observations reported here were obtained at the MMT Observatory, a joint facility of the Smithsonian Institution and the University of Arizona.

\section{Data Availability}
The scripts and notebooks used to reproduce the results and figures are available at the GitHub repository \url{https://github.com/ReshmaAnnaThomas/reaflfast_M31_FRBs_paper}. The necessary data products will be linked in the same directory or can be obtained upon request from the author. This work made use of \textsc{rfpipe} \url{https://github.com/realfastvla/rfpipe}, example \realfast localization scripts are available at \url{https://github.com/ReshmaAnnaThomas/realfast_localization}. 

\bibliography{m31frbs}{}
\bibliographystyle{aasjournal}

\end{document}